# *K*-core analysis of shear-thickening suspensions

Omer Sedes,[1] Hernan A. Makse,[2] Bulbul Chakraborty,[3] and Jeffrey F. Morris[1]

[1]*Benjamin Levich Institute and Department of Chemical Engineering, CUNY City College of New York, New York, NY 10031 USA*

[2]*Benjamin Levich Institute and Department of Physics, CUNY City College of New York, New York, NY 10031 USA*

[3]*Martin Fisher School of Physics, Brandeis University, Waltham, MA 02454 USA*

Shear thickening of suspensions is studied by discrete-particle simulation, accounting for hydrodynamic, repulsive, and contact forces. The contact forces, including friction, are activated when the imposed shear stress $\sigma$ is able to overcome the repulsive force. The simulation method captures strong continuous and discontinuous shear thickening (CST and DST) in the range of solid volume fraction $0.54 \leq \phi \leq 0.56$ studied here. This work presents characteristics of the contact force network developed in the suspension under shear. The number of frictional contacts per particle $Z$ is shown to have a one-to-one relationship with the suspension stress, and the conditions for simple percolation of frictional contacts are found to deviate strongly from those of a random network model. The stress is shown to have important correlations with topological invariant metrics of the contact network known as *k*-cores; the *k*-cores are maximal subgraphs ('clusters') in which all member particles have *k* or more frictional contacts to other members of the same subgraph. Only $k \leq 3$ is found in this work at solid volume fractions $\phi \leq 0.56$. Distinct relationships between the suspension rheology and the *k*-cores are found. One is that the stress susceptibility, defined as $\partial \sigma / \partial \dot{\gamma}$ where $\dot{\gamma}$ is the shear rate, is found to peak at the condition of onset of the 3-core, regardless of whether the system exhibits CST or DST. A second is that the stress per particle within cores of different *k* increases sharply with increase of *k* at the onset of DST; in CST, the difference is mild.

## I. INTRODUCTION

Complex fluids exhibit rheology that varies with the applied forcing. This results in behaviors seen in toothpaste or wet cement mortar, which flow only above a threshold, or yield, stress. The need for a finite yield stress to impose flow arises from the organization of suspended particles, and more precisely to the network of forces between the particles. While flow tends to destroy the force network resulting in a yield stress, with a concomitant shear thinning following yielding, roughly the opposite can also occur: flow can induce forces between particles. This is seen in shear thickening and jamming of suspensions, as hydrodynamic forces increase with shear rate and drive particles into a different state of interaction with more constrained relative motions, which has been generically termed contact.[1] The network of contact forces is closely correlated with shear thickening,[2] a behavior that has been related to shear jamming.[3,4] In this work, we consider the relationship between the network of forces between particles and the rheological properties of a shear-thickening suspension using $k$-core analysis.[5]

The network description is at an intermediate scale, between a macroscopic continuum approach and a fully-detailed microscopic particle interaction description. For granular systems, network science approaches have yielded insights into mesoscale phenomena, with particular success in providing quantitative support toward understanding of experimentally observed force chains.[6–8] The mathematical object that describes complex networks is a graph, and we use the words graph and network interchangeably. As documented in a comprehensive review by Papadopoulos et al.,[9] granular systems may be represented using different types of graphs, since the choice of the graph is not unique and is generally made based on convenience for describing the physics and geometry of the problem under investigation. The most prevalent representations are adjacency graphs,[10] which are descriptions of the contact state of the grains, and weighted graphs,[11] in which the magnitude of tangential and/or normal components of the force between contacting grains is accounted for, representing what are known as edge weights. The network can be characterized by certain metrics, including but not limited to the average network degree, the degree distribution, clustering coefficients and centrality measures.[12] Beyond these metrics, investigations of the community structure,[13,14] or topological properties like k-simplices[15] and Betti numbers[16] of these networks have proven useful in uncovering the mesoscopic dynamics of granular materials.

Jamming, a phenomenon that has been a subject of intense investigation by various methods over the last couple of decades,[17,18] has also been studied by network analysis. Jamming here means the transition of a material from a flowing to a rigid state, but maintained in the rigid state by the imposed stress. In jamming, unlike in crystallization, the material preserves its disordered structure in this solid state. The initial theoretical framework to understand this transition used $Z$, the average number of contacts a particle has with its neighbors, as the order parameter to explain the transition. The critical value of this order parameter was identified as the isostatic point, defined as the minimum possible number of contacts required for mechanical stability based on the Maxwell criteria for rigidity. For granular packings of frictionless spheres, this number is twice the number of dimensions, $Z_{iso} = 2d$, and this mean-field scalar quantity was shown to be a good predictor of the jamming point.[19,20]

For frictional systems, jamming has been observed at conditions ranging from $d+1 < Z_{iso} < 2d$ in simulations and experiments,[21,22] with the value of $Z_{iso}$ decreasing as the interparticle friction coefficient increases. Given the lack of a clear isostatic point, several approaches to jamming of frictional packings have been applied. Initially, a generalization of isostaticity to include the effect of friction was proposed.[23,24] A number of further investigations focused on the emergence, propagation, and percolation of structural rigidity within the contact network.[25] It was discovered in studies of tree-like networks that connectivity percolation and rigidity percolation were not the same. Connectivity percolation was a necessary but not a sufficient condition for rigidity percolation[26] as shown for granular materials by Feng,[27] where contact connectivity percolation was realized at $Z$ values below the isostatic condition.

In order to identify rigid clusters, a 'pebble game' algorithm which keeps track of the constraints based on Laman's theorem was used for two-dimensional packings of particles.[28] Schwarz and co-workers[29] extended the pebble game beyond Laman's minimal rigidity criterion and explored local versus global rigidity in a network, the size distribution of rigid clusters, and the role of spatial correlations for frictionless granular packings.[30–32] Henkes et al.[33] applied an extended pebble game algorithm to frictional jamming, successfully identifying floppy and rigid clusters and demonstrating the percolation of rigid structures at the jamming transition for simulations of two-dimensional packings. Beyond granular systems, the rigidity percolation has been shown to occur in colloidal gels as well, by Zhang et al.[34] using molecular dynamics simulations. Extension of this algorithm to rigidity percolation in three dimensions for frictional packings has not been developed, although the recently introduced rigid graph compression algorithm appears to be a promising method.[35]

Beyond two dimensions, rigidity percolation has been considered on Cayley trees by Mourkazel et al.[36] These authors established an equivalence between rigidity percolation and what was then termed bootstrap percolation,[37] and currently known as *k*-core percolation.[38] The *k*-core is a topological invariant of a graph defined[39] as the maximal subgraph where each site (or node) has at least degree *k*. Similarity to rigid clusters is seen in *k*-core clusters for $k \geq 3$, as these display non-local effects. For example, removal of a bond may trigger the collapse of a single cluster to a large number of smaller clusters[38,40] involving sites far from the removal site. For simple connectivity percolation, removal of a bond would at most split a large cluster into two at the location of the bond removal.

Correspondence of the geometric nature of dynamical arrest with *k*-core percolation was shown by Selitto et al.[41] in study of glassy dynamics in the Fredrickson-Andersen model[42] on Bethe lattices. Following this work, Schwarz *et al.*[29] argued that the jamming transition is analogous to a *k*-core transition, emphasizing that the Bethe lattice *k*-core percolation is a critical phenomenon. They demonstrated a susceptibility and a correlation length of the *k*-core transition and found agreement of their critical exponents with those of simulations of particle packings near the jamming transition. The authors noted, however, that *k*-core models lack the mechanical constraints associated with jamming.

More recently, Morone et al.[43] determined *k*-cores of the contact networks in sheared three dimensional simulations of dry frictional packings near their jamming point. In addition to observing a discontinuous jump of shear stress near the isostatic point, they found a precursor increase in shear stress as the isostatic point is approached; this increase was found to be roughly coincident with the appearance of the 3-core. Furthermore, they demonstrated that the appearance and size of the *k*-cores in the contact network corresponded to the values predicted in random graph theory by Pittel et al.[44]

Here, we consider shear-thickening suspensions. Even simple suspensions, composed of near-hard spheres in a Newtonian liquid, exhibit what is known as discontinuous shear thickening (DST) at sufficiently high solid fraction, $\phi$. In DST, the viscosity rises discontinuously with an increase in the imposed shear rate. At lower $\phi$, the same suspension undergoes a weaker or continuous shear thickening (CST), while at higher $\phi$ it may undergo shear jamming.[45] This behavior has been related to the development of stress-induced contact frictional interactions; the threshold stress for onset of these contacts is assumed to scale with a repulsive force between particles.[1,46–48] Thus, beyond the potential relation to the jamming transition and to rigidity, a further motivation to investigate the *k*-core structure of contact networks in suspensions is related to the concept that a threshold stress is required to drive a change in the microscopic interactions. The physical basis for the scaling of this threshold stress has received attention.[49,50] Morone et al.[5] analyzed the collapse of mutualistic ecosystems, demonstrating that a sigmoidal interaction with a threshold and saturation parameter would lead to a *k*-core 'tipping point.' They supported this claim by considering models of neural and gene regulatory networks with such interactions, and found similar relations between the *k*-cores and the tipping points of these dynamical systems.

In models of dense suspensions that exhibit shear thickening, such a sigmoidal form of interaction with a threshold has been applied. This models the transition (with increasing applied stress) in dominant stress generation mechanism from hydrodynamic forces to frictional interparticle forces. This is able to rationalize the salient features of the rheology[45–47,51,52]. This prompts an investigation of whether the fixed points associated with this shear thickening transition also relate to *k*-core development in the frictional contact network.

For dense suspensions, study of the contact network due to the formation of enduring contacts created during the shear-thickening transition has been limited. Mari et al.[47] showed qualitatively that frictional contacts percolate in all directions in the thickened state while they are very sparse prior to shear-thickening. Boromand et al.[53] analyzed the contact network of shear thickening suspensions simulated by the dissipative particle dynamics method, allowing frictional interactions between particles. This study described size and radius of gyration of clusters of particles connected by frictional contacts, and the authors argued that the formation of a giant connected component led to DST, although some of us have argued this relationship is not completely accurate in more recent work[54]. In a recent study applying a different network analysis, Gameiro et al.[55] extracted topological features of the contact networks of suspensions near the shear-thickening transition by methods of persistent homology, a force-thresholding analysis. They concluded that the loop structure growth within the contact network has the clearest correlation with the stress-induced viscosity increase.

Here, we seek a more thorough understanding of the shear-induced contact force network and its relation to the shear-thickening transition, by which we mean the transition from dominance of lubrication hydrodynamics to frictional contact forces at the microscale that is associated with sharp increases in the viscosity and normal stress of the bulk suspension.[1] We begin in the following section by describing briefly the simulation method, which involves a stress-induced transition from lubricated to frictional interactions, as developed by Mari et al.[47] We follow with the characterization of the network through time averages and temporal distributions of number of frictionally-interacting particles and the degree distributions of the stress-induced frictional contact network (FCN). We proceed to consider the growth of the FCN, and determine the conditions resulting in formation of a giant connected component (GCC), i.e. a percolated structure with connections defined here by frictional contacts. Next, we investigate the *k*-core organization, relating the network structure to the bulk rheological properties. An essential finding is that the condition of maximum variation of stress with shear rate for a given solid fraction corresponds to the emergence of the 3-core in the FCN. Finally, we investigate the distribution of stress on the *k*-core structures, showing a direct relationship between the stress-bearing 3-cores and the shear-thickened state under DST.

## II. SIMULATIONS

We use the lubricated flow-discrete element method (LF-DEM)[47,56] to simulate dense noncolloidal suspensions in a narrow range of $\phi$, considering $\phi = 0.54, 0.55, 0.555$ and $0.56$. For the interparticle forces and bidispersity simulated here, this range of $\phi$ spans the shear thickening transition, with $\phi = 0.54$ and $0.56$ clearly exhibiting CST and DST, respectively.

The simulation method considers spherical particles immersed in a Newtonian liquid that lubricates the particle surfaces but neglects long-range hydrodynamic interactions. Motivated by the concept of 'lubrication breakdown,' i.e. the failure of hydrodynamic lubrication to maintain finite surface separation in sheared dense suspensions, as shown by Ball and Melrose,[57] a key feature is that the method allows for frictional contact interactions between particles when the imposed shearing force overwhelms a repulsive inter-particle force.

The particles are bidisperse, with radii $a_1 = a$ and $a_2 = 1.4a$, to avoid ordering observed in dense monodisperse suspensions.[58,59] Half of the total solid volume is contributed by particles of each radius. The $N = 500$ particles are confined to a cubic unit cell of volume $V$ fixed by the desired value of $\phi$. The unit cell is periodically replicated in all three directions and sheared according to Lees-Edwards boundary conditions. The simulations reported were performed at an imposed shear rate, $\dot{\gamma}$, with the shear stress, $\sigma(t)$, fluctuating as the suspension samples configurations. The shear flow is $u_x = \dot{\gamma} z$ implying flow, gradient, and vorticity directions are $x$, $z$, and $y$, respectively.

We consider over-damped motion (conditions of zero inertia), so that each particle satisfies a force balance between finite-range hydrodynamic ($\mathbf{F}_H$) and conservative forces ($\mathbf{F}_R$), as well as contact forces ($\mathbf{F}_C$). Here, we use what is termed the critical load model[47] to capture the influence of the repulsive forces: in this case we do not have finite-range repulsion, but instead impose a threshold compressive contact normal force of magnitude denoted $F_R$ to indicate its role in replacing the repulsion, above which friction is activated in the contact force; for normal forces less than $F_R$, a frictionless contact is modeled. The particles thus obey the simple force balance

$$0 = \mathbf{F}_H + \mathbf{F}_C, \tag{1}$$

along with a similar balance of hydrodynamic and frictional torques.

The hydrodynamic force $\mathbf{F}_H$ accounts for single-particle Stokes drag and pair-particle lubrication, which is associated with the fluid in the zone of closest approach of neighboring particle surfaces and is the dominant pair hydrodynamic interaction for large solid fraction.[47] We write the lubrication in terms of the velocities of two particles, $\mathbf{U}_1$ and $\mathbf{U}_2$, as $\mathbf{F}_H = -\mathbf{R}_2(h) \cdot (\mathbf{U}_2 - \mathbf{U}_1)$, with the pair resistance tensor $\mathbf{R}_2 \sim (h+\delta)^{-1}$ for the motion along line of centers, where $h = r - (a_1 + a_2)$, $r$ is the center separation of the pair, and here $\delta = 10^{-3}$. Including $\delta$ yields a finite lubrication resistance at contact ($h = 0$), and can be considered as representative of a roughness lengthscale, but with no further physical modeling of roughness. When contact occurs, the contact force, $\mathbf{F}_C$, comes into play. When the interparticle normal force exceeds the 'critical load' value $F_R$, friction is activated, and satisfies the Coulomb criterion, $F_{C,t} \leq \mu F_{C,n}$, relating tangential ($F_{C,t}$) and normal ($F_{C,n}$) components. Here, we study only $\mu = 1$. Prior work has shown that varying $\mu$ alters the frictional jamming fraction and the volume fraction for DST,[47,48] but does not change the qualitative physics.

Each simulation was run to a strain of 20. The simulation quantities of interest for this study are the individual force moment contributions to the stress (often called stresslets), the network of interactions, and the bulk rheological properties of viscosity and normal stresses. These outputs are sampled at strain intervals of 0.01, yielding $T = 2000$ samples per simulation.

## III. NETWORK REPRESENTATION OF FRICTIONAL CONTACTS

We represent the frictional contact state of the suspension using a network (or a graph), where the particle centers are the nodes (vertices) and the frictional contacts form the edges. The adjacency matrix corresponding to this undirected network, $\mathbf{A}$, is $N \times N$ and symmetric, where $N$ is the number of particles in the simulation; $A_{ij} = 1$ if there is a frictional contact between particles $i$ and $j$, and otherwise $A_{ij} = 0$. The degree or coordination number of a particle is defined as $Z_i = \sum_{j=1}^{N} A_{ij}$ and it gives the number of frictional contacts in which particle $i$ participates; this is evaluated at each sampling and statistics are determined from all samples. The average coordination number, or equivalently the mean degree of the network, can also be directly calculated from the adjacency matrix by $Z = (1/N) \sum_{i=1}^{N} Z_i = (1/N) \sum_{i=1}^{N} \sum_{j=1}^{N} A_{ij}$. It is important to note that the FCN, and hence the adjacency matrix corresponding to this network, is not static, meaning we have an evolving network as frictional contacts are formed or broken as the material flows.

At each sampling point, we record the bulk rheological properties, the individual particle contributions to these properties, and the corresponding adjacency matrix. This simultaneously gathered information allows us to map the frictional network properties to the rheological state of the suspension. We are interested in the long-time average, represented by $\langle \rangle$ and assumed to equate to an ensemble average, as well as the temporal distribution of network metrics discussed in subsequent sections.

## A. Degree Distribution and Average Network Size

We consider first the degree distribution of the frictional contact network. We calculate $P(Z_i)$, the probability of particle $i$ having degree $Z_i$, by considering all the samplings of the simulations at each shear rate. We construct a histogram of particles of degree $Z_i$, then normalize the counts by $N \times T$, the product of the number of particles $N$ and samplings $T$. In all further work here, $Z$ refers to a number of frictional contacts and does not account for frictionless contacts.

The resulting degree distributions are shown in Fig. 1 for $\phi = 0.54$ and $\phi = 0.56$, corresponding to CST and DST, respectively. Note that the shear rate is normalized by $F_R/(6\pi\eta a^2)$. For both volume fractions, at the distributions corresponding to the lower shear rates the particles have few frictional contacts, and thus there is high probability value for $Z_i = 0$. For $\phi = 0.54$ at the lowest shear rate displayed, $\dot\gamma = 0.03$, approximately 10% of the particles have either one or two contacts, and for $\phi = 0.56$ particles with frictional contacts are less than 5% of the total for the lowest shear rate, $\dot\gamma = 0.015$. With increasing shear rate, $P(Z_i = 0)$ decreases and the contact network forms as $P(Z_i > 0)$ values become nonzero. The value of $Z_i$ with the highest probability, i.e. the mode of the distribution for the values of $Z_i > 0$, shifts to higher values of $Z_i$ at elevated shear rate. Fig. 2 shows that at the largest shear rates shown both the mean frictional contact number and relative viscosity have largely saturated, but in data not shown, we do find that the peak shifts to $Z_i = 4$ at $\dot{\gamma} > 0.12$ for $\phi = 0.54$. The distribution of $Z_i$ values higher than the mode resembles an exponential decay towards $Z_i = 8$, which is the largest value observed for all $\dot{\gamma}$ and $\phi$ considered in this study.

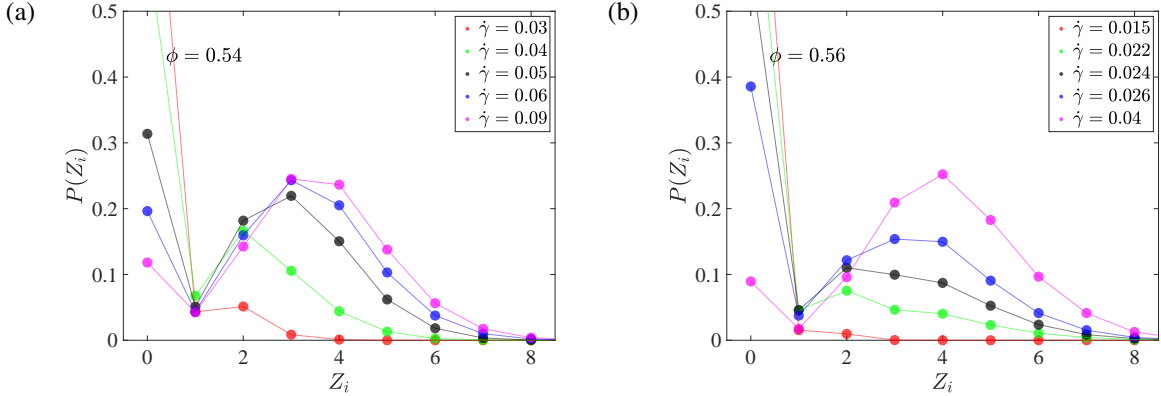

FIG. 1: Degree distributions $P(Z_i)$ for (a) $\phi = 0.54$ and (b) $\phi = 0.56$, showing the distribution of frictional contacts per particle $Z_i$ for different $\dot{\gamma}$.

The growth of the FCN as a function of the shear rate can be readily visualized by considering the fraction of total particles in frictional contact with one or more other particles, $n$. In the top panel of Fig. 2, the fraction of particles in frictional contact, determined by summing the number of particles with degree $Z_i > 0$ and dividing by the number of particles $N$, is shown as its time average $\langle n \rangle$ as a function of $\dot{\gamma}$. Comparing $\langle n \rangle$ to the development of suspension relative viscosity (i.e., apparent viscosity normalized by the suspending fluid viscosity) in the bottom panel reveals the correlation of contact network growth with the rheological transition. Note that particles enter and leave the network as the shearing takes place, and that $\langle n \rangle < 1$ even at the largest $\dot{\gamma}$ studied. Thus, there are always particles that are not a part of the frictional contact network.

The time average of the mean degree, $\langle Z \rangle$, is shown as a function of shear rate in the middle panel of Fig. 2. Similar to $\langle n \rangle$, the growth of $\langle Z \rangle$ follows that of the viscosity, and saturates as $\dot{\gamma} \to \infty$, with increasing saturation values for larger $\phi$.

We see that the FCN is an emergent object. For the low-stress state or $\dot{\gamma} \to 0$, the viscosity is non-zero and increases with $\phi$, whereas the frictional contact network is non-existent regardless of the value of $\phi$ as $\dot{\gamma} \to 0$, since $\langle n \rangle = 0$ and $\langle Z \rangle = 0$ at sufficiently small stress. The formation of the network begins when the suspension is sheared sufficiently rapidly, thus generating sufficient stress to drive particles into the frictional state. This is in contrast with dry granular materials, where a FCN may exist under static, zero deformation conditions.

Figure 3 displays the distributions for the fraction of particles in the frictional network, $n$. In

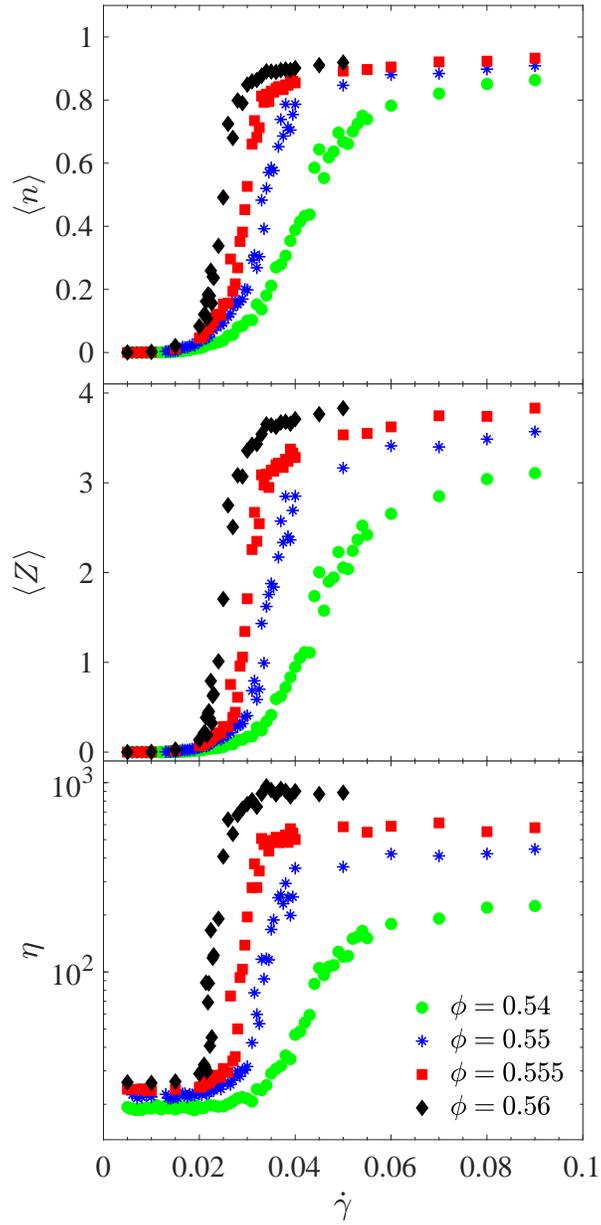

FIG. 2: The growth of the frictional contact network as a function of the shear rate, in comparison with suspension viscosity. (top) Time averaged fraction of particles that are part of the frictional network, $n$; (middle) time averaged network degree, $\langle Z \rangle$; and (bottom) suspension relative viscosity, $\eta$.

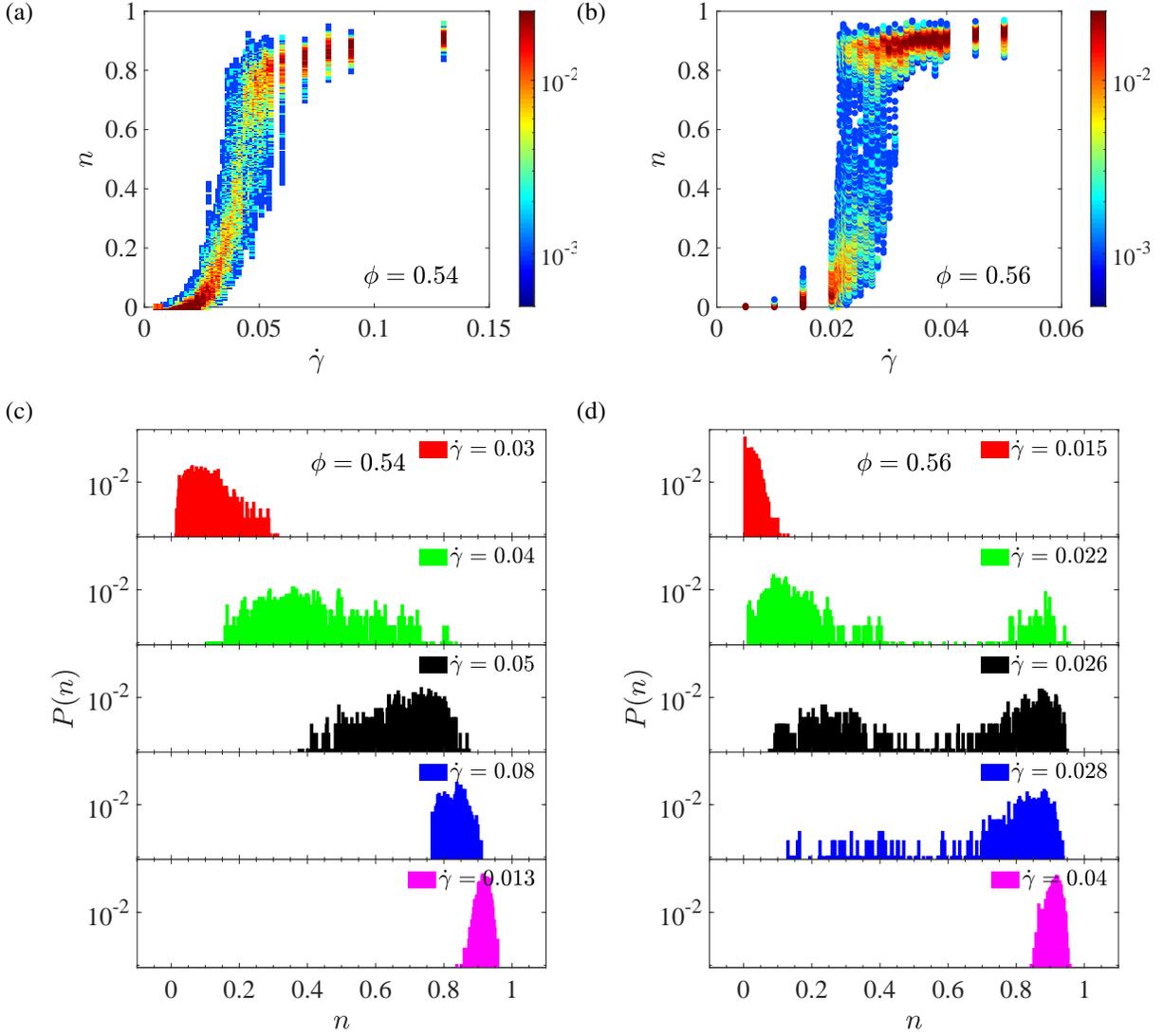

FIG. 3: Temporal distribution of the fraction of particles in the frictional network (i.e., in frictional contact with one or more other particles), $n$, for $\phi = 0.54$ (a, c) and $\phi = 0.56$ (b, d). In (a, b), evolution of $n$ with $\dot{\gamma}$ is, where $P(n)$ of each value of $n$ is represented by its color. In (c, d) distributions $P(n)$ are shown for selected values of shear rate $\dot{\gamma}$.

Fig. 3a and 3b, observed values of $n$ at several shear rates are plotted for $\phi = 0.54$ and $\phi = 0.56$, with the color of a data point indicating the probability of finding the value of $n$. Fig. 3c and Fig. 3d display distributions of the same quantities for selected shear rates, in $P(n)$ format. We see a difference in the form of the distributions in the transition region for CST and DST. For $\phi = 0.54$, under CST conditions, the distribution has a single maximum value for every $\dot{\gamma}$. By contrast, for $\phi = 0.56$, under DST conditions, $P(n)$ is bimodal in the transitional region $0.02 < \dot{\gamma} < 0.033$, as seen at $\dot{\gamma} = 0.024$ and $0.026$ in Fig. 4. Similar bimodal stress distributions are seen in prior work.[54]

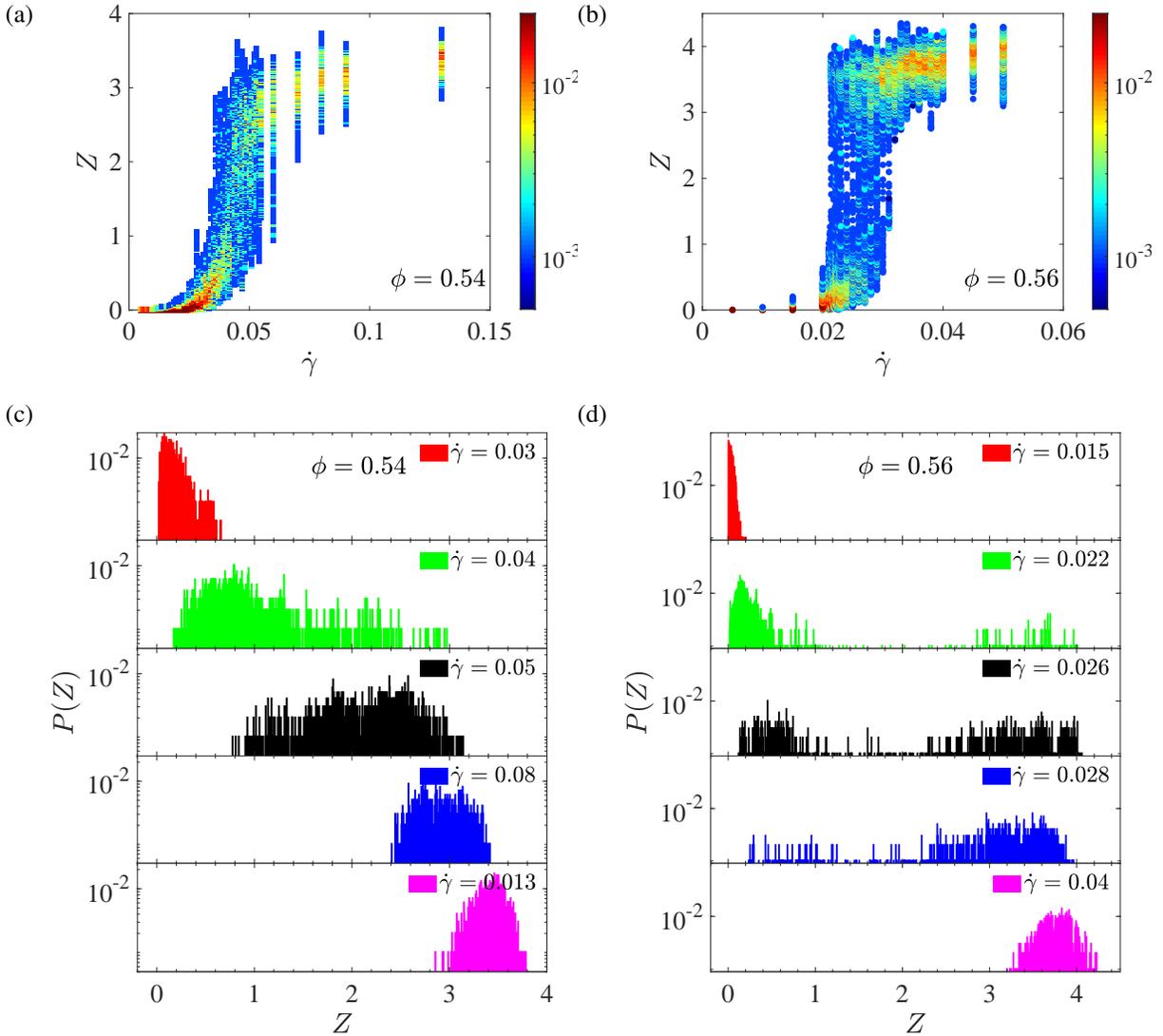

FIG. 4: Temporal distribution of the frictional contact network degree, $Z$, as a function of shear rate, $\dot{\gamma}$ for $\phi = 0.54$ and $\phi = 0.56$. The top plots (a, b) display the evolution of $Z$ as a function of $\dot{\gamma}$, where $P(Z)$ of each value of $Z$ is represented by its color. The bottom plots (c, d) show some of the $Z$ distributions $P(Z)$ for different values of $\dot{\gamma}$.

In Fig. 4 we present the temporal distribution of the degree, averaged at each sampling instant over all particles in the simulation unit cell, and denoted as $Z$. The DST condition again displays temporal coexistence of two values. A striking feature is that at $\phi = 0.56$, a jump from $Z \approx 0.5$ to $Z \approx 3$ occurs over a narrow range of $0.022 \leq \dot{\gamma} \leq 0.028$ where DST takes place, indicating the suspension may have a wide range of instantaneous mean contact states at the same shear rate.

The implication of Fig. 4 is that structures of higher connectivity can form and collapse under the same shear rate in a short time or small strain. These distributions reflect the temporal stress distributions discussed in Sedes et al.,[54] and show a correspondence between the stress exerted by the suspension and the number of particles in the contact force network. When $Z$ is plotted against the instantaneous suspension shear stress, $\bar{\sigma}$, in Fig. 5, it is clear that the frictional contact number is a monotonically increasing function of $\bar{\sigma}$ for any $\phi$, with only modest variation for the $\phi$ values studied here, despite the fact that these span the CST-DST boundary. This one-to-one correspondence of mean contact number and bulk shear stress is significant, as it provides a mapping between a fundamental property of the FCN and the rheology of the suspension, enabling us to connect the transitions in the topology of the network to changes in suspension rheology. Note that this result of monotonic $Z(\bar{\sigma})$ implies that divergence of $\partial\sigma/\partial\dot{\gamma}$ or discontinuity of $\sigma$ at DST also applies to $\partial Z/\partial\dot{\gamma}$.

We further characterize the network by the net degree, i.e., the number of contacts per particle restricted to those particles in the FCN. This measure removes the unconnected particles, or 'rattlers.' For this purpose, it is only necessary to determine the number of edges $Z$ normalized by the fraction of nodes $n$, and the quantity $\langle Z_{net} \rangle$ is shown in Fig. 6. While $\langle n \rangle$ and $\langle Z \rangle$ follow simple sigmoidal trends, $\langle Z_{net} \rangle$ exhibits more distinct regimes. The $\langle Z_{net} \rangle$ curves for all $\phi$ values studied nearly overlap up to $\dot{\gamma} \approx 0.02$ where $\langle Z_{net} \rangle \approx 1$. At this point, the various $\phi$ curves separate, with $\langle Z_{net} \rangle \approx 2$ reached at smaller $\dot{\gamma}$ with increasing $\phi$. The rate of growth, $\partial \langle Z_{net} \rangle / \partial \dot{\gamma}$, for $0 < \langle Z_{net} \rangle < 1$ is higher than the rate of growth in the range $1 < \langle Z_{net} \rangle < 2$ for all $\phi$. In the third regime, from $\langle Z_{net} \rangle \approx 2$ to its saturation value, the disparity between different volume fractions is more pronounced, and $\partial \langle Z_{net} \rangle / \partial \dot{\gamma}$ is larger in this regime compared to the second, and is similar to the first where $0 < \langle Z_{net} \rangle < 1$. The intermediate regime, where the rate of contact growth with increase of shear rate slows, appears to end coincident with onset of the most rapid rise in the viscosity.

## B. Percolation and the Emergence of a Giant Connected Component

It is of interest in study of networks to identify whether a giant component exists, and if so under what conditions[60,61]. Mendes et al.[62] note that the discovery and the description of this structural transition initiated the study of random networks. In random networks, there exists a critical probability, $p_c$, where a percolation transition occurs; $p$ is the probability that two nodes selected at random have an edge between them. For $p > p_c$, a giant cluster that spans the entire network exists, whereas for $p < p_c$, there exist many isolated, smaller clusters.[63]

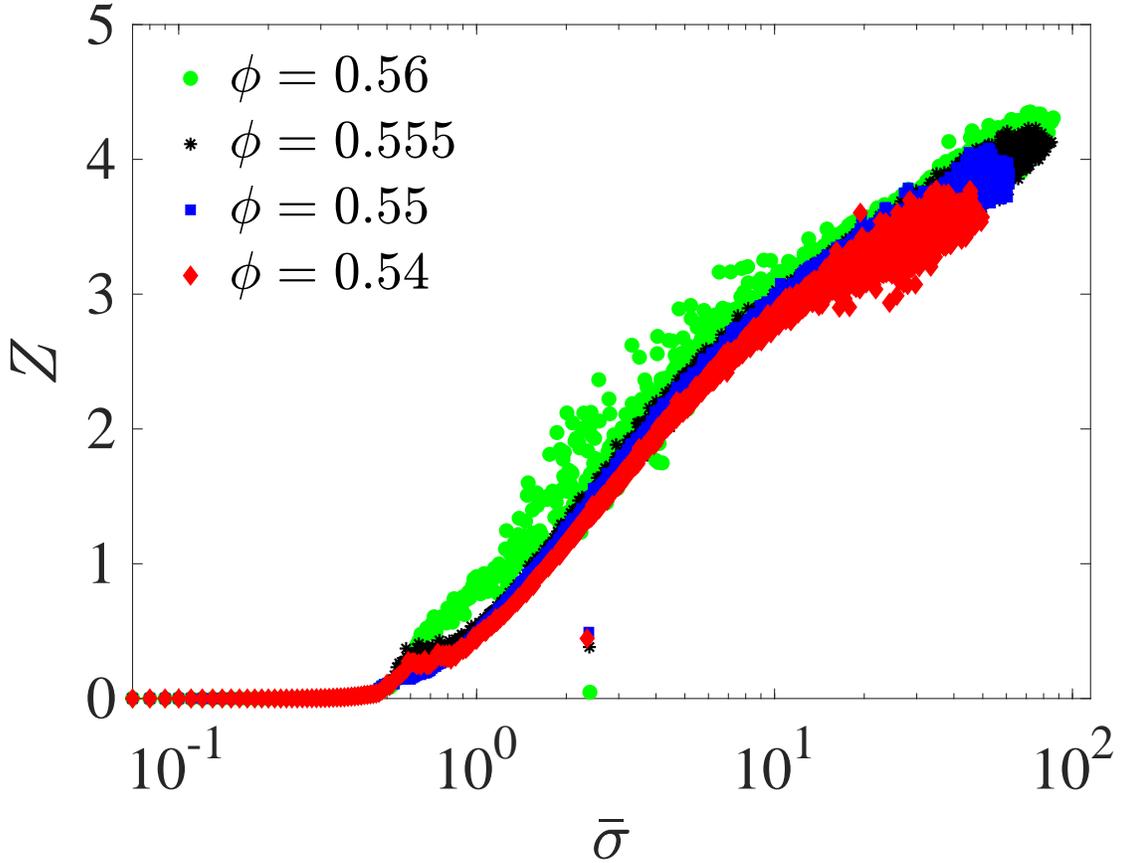

FIG. 5: Frictional contact network degree Z, as a function of suspension shear stress, $\bar{\sigma}$ for $\phi = 0.54, 0.55, 0.555$ and $0.56$. Each point represents a time sample, and the full range of shear rate $\dot{\gamma}$ is represented. The change in slope at $Z \approx 0.28$ is shown later to correspond to contact percolation.

For the FCN of shear thickening suspensions simulated here, we seek to determine if a topological transition related to the birth of giant (or percolating) connected component (as noted, a GCC) exists. To do so, we identify the largest connected component (LCC) of the frictional contact network at every sampling instant.

In Fig. 7a, we plot the time-averaged fraction of the system total number of particles in the LCC, $\langle f \rangle$, as a function of $\dot{\gamma}$. This is displayed for $\phi = 0.54, 0.55$ and $0.56$, and in each case, the mean LCC size follows a sigmoidal curve from zero for low shear rates to a saturation value of $\langle f \rangle \approx 0.9$ at large $\dot{\gamma}$. Normalizing the size of the LCC by the number of frictional particles, and defining $f_{net} = (N/n)f$, we show in Fig. 7b that the saturation value of $\langle f_{net} \rangle$ approaches unity, meaning that essentially all of the frictional particles become a part of the LCC, i.e. at saturation a single GCC typically spans the FCN. Interestingly, the emergence of this giant component is not a criterion that delineates between DST and CST, since it is observed to occur for $\phi$ that are in each of these shear-thickening regimes. We conclude that the emergence of a GCC does not have a clear connection to the rheological transition of the suspension.

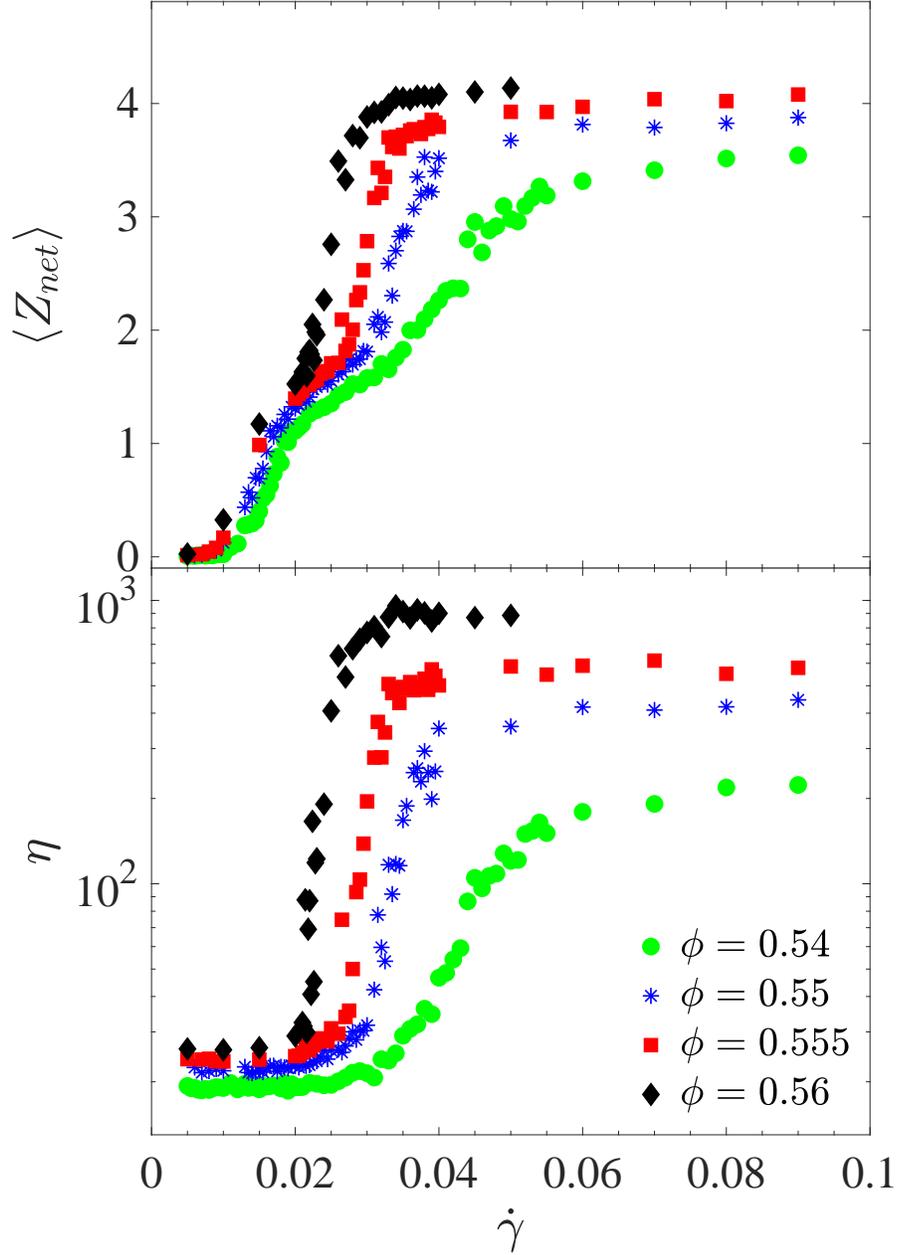

FIG. 6: Time averaged net degree (top), $\langle Z_{net} \rangle$, and corresponding suspension relative viscosity (bottom), both as a function of shear rate, $\dot{\gamma}$, for $\phi = 0.54, 0.55, 0.555$ and $0.56$.

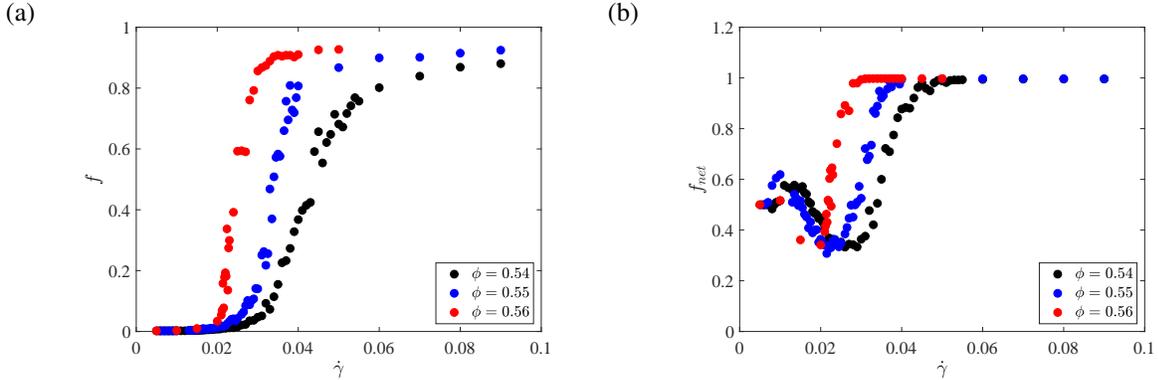

FIG. 7: The fraction of total particles in the largest connected component, as a function of shear rate, $\dot{\gamma}$, for $\phi = 0.54$ (black), $\phi = 0.55$ (blue), $\phi = 0.56$ (red), (a) normalized by total number of particles, $N$, (b) normalized by the number of particles participating in frictional contacts.

We plot in Fig. 8 the linear extent of the cluster in all three directions normalized by the simulation box size, along with $f$, the fraction of particles in the LCC. For all $\phi$ studied here, the length of the cluster reaches the box length first along the gradient direction ($z$), followed closely by the flow ($x$), and finally the vorticity ($y$) direction. At the shear rate of initial percolation of the LCC in all directions, $\langle f \rangle \approx 0.6$, well below its saturation value. Thus, the GCC continues to grow in number of particles for $\dot{\gamma}$ increasing above the value associated with cluster percolation.

In random networks, the emergence of a giant component is a function of $Z$, and there exists a $Z_c$ that marks the emergence of the giant component.[64] In section III A, we showed that the FCN degree $Z$ can be directly mapped to the stress. We now probe how the mean degree and the size of the LCC are correlated, specifically seeking to identify $Z_c$ for the onset of a GCC in the FCN. We plot $f$ and $\dot{\gamma}$ as a function of network degree, $Z$, for all volume fractions under consideration in Fig. 9. Considering the size of the LCC, at all volume fractions we observe a brief linear regime in the range of $Z$ between 0 and 0.3, followed by a sharp increase which in fact marks the emergence of the GCC. A precise method to identify $Z_c$ in uncorrelated networks is provided by the Molloy-Reed criterion[65] that the emergence of the giant component occurs when the mean number of next-nearest neighbors, $\sum Z_i^2 - Z$, exceeds the mean number of nearest neighbors, $Z$. Based on this, a susceptibility measure for the network can be defined as $s \equiv 1 + \sum Z_i^2 / (2Z - \sum Z_i^2)$, such that the divergence of *s* corresponds to the emergence of the GCC in the suspension. We plot *s* as a function of the network degree $Z$ in Fig. 9. Identifying $Z_c$ as the point of divergence of *s*, we find this corresponds to the transition to a steeper slope in the curve that shows the size of LCC, $f(Z)$. It is significant that the mean degree at this transition point is the same across all volume fractions, and is identified as $Z_c \approx 0.28$.

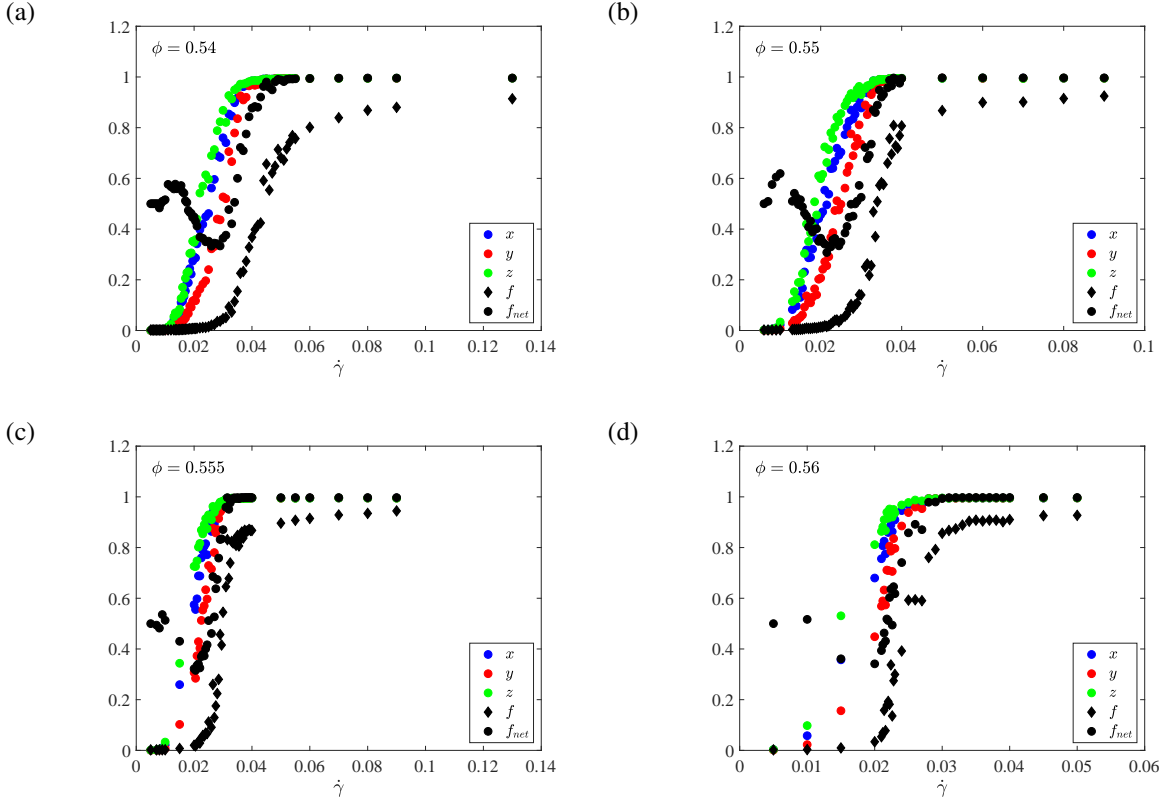

FIG. 8: The mean size of the largest connected component, as a function of shear rate, in number of particles (black), its length in *x* (blue), *y* (red) and *z* (green) dimensions (plotted as the fraction of the unit cell side length), for (a) $\phi = 0.54$, (b) $\phi = 0.55$, (c) $\phi = 0.555$, and (d) $\phi = 0.56$.

From a network theory perspective, a noteworthy result from this section is that the FCN deviates significantly from a random graph. The Erdos-Renyi random graph model predicts[64] the birth of a giant connected component at mean $Z_{c,random} = 1$, whereas we find $Z_c \approx 0.28$ for the FCN in a sheared suspension. The largeness of the difference is notable. We further note that divergence of the Molloy-Reed susceptibility, *s*, does not correspond to the maximum or divergence (when singular) of either $\partial Z/\partial \dot{\gamma}$ or $\partial \sigma/\partial \dot{\gamma}$. The emergence of a GCC is found instead to roughly correspond to the onset (with respect to $\dot{\gamma}$) of the shear thickening transition.

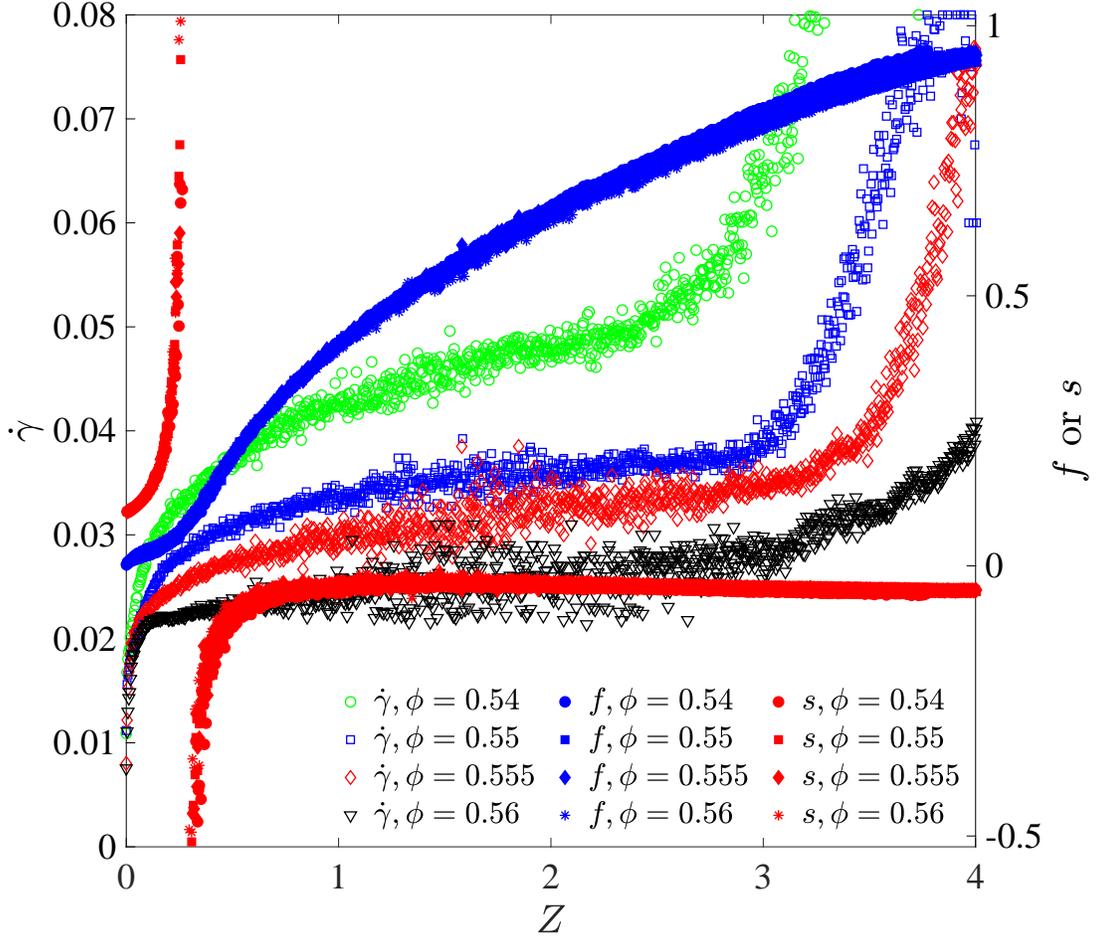

FIG. 9: The size of the largest connected component, $f$, the Molloy-Reed susceptibility, $s$, and the shear rate, $\dot{\gamma}$ (open symbols), as a function of the network degree, $Z$, for all simulated volume fractions. Note that the curves of $f$ and $s$ collapse for all $\phi$, with $f$ showing a change of slope at the divergence of $s$ for $Z \approx 0.28$.

We find that neither local measures such as the mean degree (contact number) nor the simple long-range connectivity of percolation is a clear indicator of the mechanical property changes seen in the DST transition. Structural transitions of higher connectivity than simple contact percolation must be investigated to link the topological changes in the frictional network directly to the mechanical basis for onset of DST, and for this we turn to *k*-core analysis.

## IV. *k*-CORES OF THE FRICTIONAL CONTACT NETWORK

A *k*-core of a graph is defined[66] as a maximal subgraph where each vertex of this subgraph has degree at least *k*. This implies any member of the set of particles associated with this sub-network has at least *k* contacts with other members of the same set.

We determine the *k*-core by a pruning process. First, we remove all nodes (the vertices) of a network (a graph) with degree less than *k*, and also remove the edges connected to these nodes. Now, we repeat this step until no nodes with degree less than *k* are left. At the end, we are left with a graph where the degree of each node is greater than or equal to *k*. We note that, in general, this graph can be disconnected.

*Cores and shells:* Upon completion of this procedure for all $k \leq k_{max}$, one is left with a nested structure of cores, such that particles that belong to a specific *k*-core also belong to all the cores with smaller *k*, from $k-1$ to 0. This can be reasoned upon considering an example in which $k = 3$, where all the particles belonging to this core all have $Z_i \geq 3$; therefore each particle automatically satisfies the condition of having a minimum degree of 2 or 1, making all the 3-core particles also members of the 2 and 1-cores. If one considers the reverse statement, whether a particle belonging to *k*-core belongs to the $(k+1)$-core, this is not true for every particle in the *k*-core, and the particles that belong to the *k*-core but not to the $(k+1)$-core form the *k*-shells of the connected component.

For random graphs, the emergence of *k*-cores with $k \geq 3$ was studied by Pittel et al.,[44] who showed that a giant *k*-core for $k \geq 3$ appears suddenly when the mean network degree reaches a threshold, $Z_{c,k}$. For $k = 3$, they showed that the emergence of a 3-core is at $Z_{c,3} \doteq 3.35$. The size of the 3-core in the random network emerges, as a fraction of number of vertices, at $f_{3,c} \doteq 0.27$ and grows with further increase of *Z*.

For a packing of monodisperse frictionless disks in two dimensions,[29] the maximum degree is given by $Z_{max} = 6$, and for monodisperse spheres in three dimensions, $Z_{max} = 12$. Our system has slight bidispersity so the maximum *Z* value differs slightly and is not precisely known, but nonetheless is expected to be similar to the value of 12 for 3D. In section III A, we saw that for shear-thickening suspensions, the maximum mean degree found was significantly lower at $Z = 8$. We turn now to the *k*-core analysis.

We begin our investigation of the *k*-core and *k*-shell structure of the frictional contact networks by a decomposition of the largest component. The sizes of the *k*-core and *k*-shell structures belong-ing to the LCC, identified at each sampling instant, are determined and then related to the rheology. This is done for both CST and DST conditions. Following this, the largest $k$-core to which a particle belongs within the LCC (the particle's 'coreness' or $k$-shell) is determined. Figs. 10 and 11 provide visualizations of the clusters that are found.

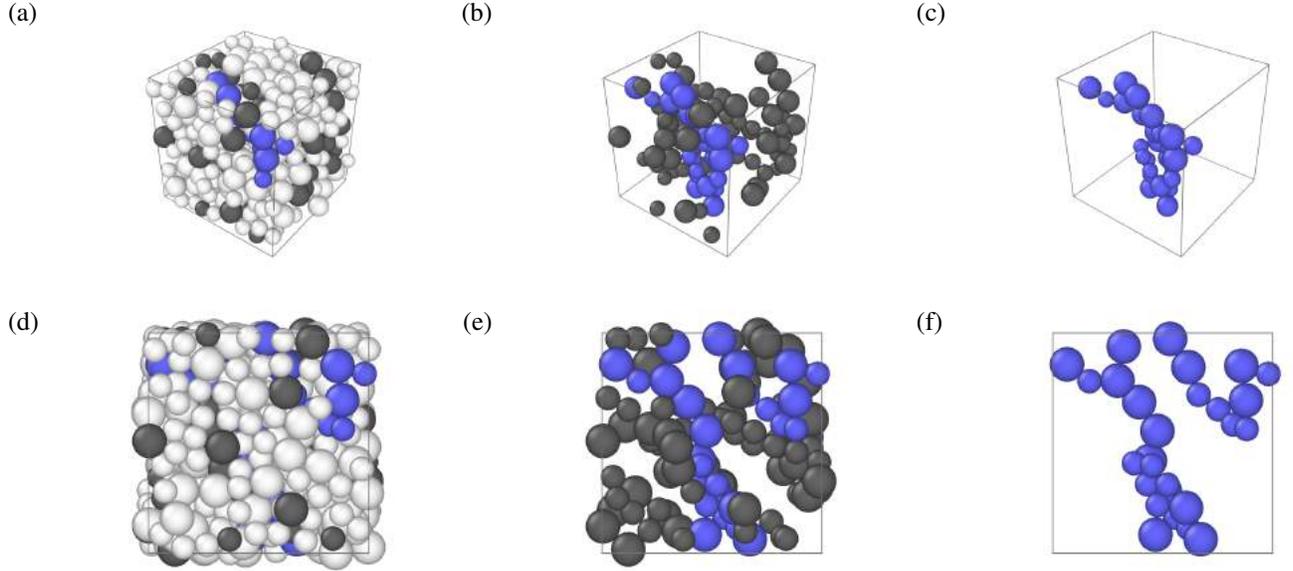

FIG. 10: A snapshot of the simulation for $\phi = 0.55$, $\dot{\gamma} = 0.025$. All particles visualized on (a, d), 1-shell (dark gray shading) and 2-core particles are visualized on (b, e), and only 2-core particles (blue shading) are visualized on (c, f). Unshaded particles are not in the frictional network.

The number $N_k$ of particles for each coreness $k$ was normalized as $f_k = N_k/N$, with $N$ the total number of particles. These operations yield the size of each $k$-shell as a fraction of the system size. The $f_k$ values for all $\phi$ as a function of shear rate are shown in Fig. 12.

From Fig. 12, we see that the maximum coreness observed is $k = 3$. This is significantly less than the upper limit established by the largest value in the degree distributions, i.e. $Z_i = 8$ as shown in Fig. 1: within the 3-cores lie particles that have substantially more than three frictional contacts. A further observation is that for all $\phi$, 1- and 2-cores emerge in a continuous fashion. The 1-cores emerge first, and as the shear rate becomes larger, 2-cores emerge, relegating the particles with coreness 1 to 1-shell status. By contrast, the 3-cores emerge discontinuously, suddenly jumping at a specific shear rate from 0 to a finite fraction of the particles, in agreement with the theoretical results of Pittel et al.[44] for the sudden emergence of the 3-core in a random graph.

For all $\phi$, there are shear rates for which both $k = 2$ and $k = 3$ structures are observed, with $f_2$

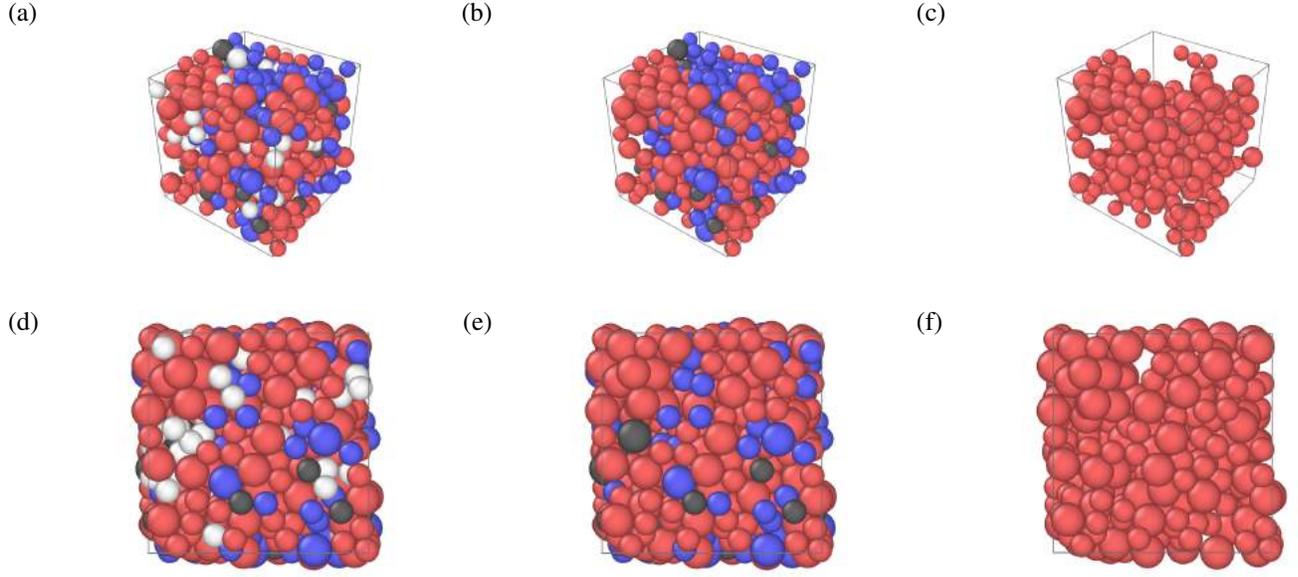

FIG. 11: A snapshot of the simulation for $\phi = 0.55$, $\dot{\gamma} = 0.0375$. All particles visualized on (a, d), 1 (black shading) and 2-shell particles (blue shading) along with 3-core particles (red shading) are visualized on (b, e), and only 3-core particles are visualized on (c, f). Unshaded particles in (a,d) are not in the frictional network.

and $f_3$ values fluctuating. Under these conditions, the $f_3$ values are either near 0 or distributed in a range between 0.2 and 0.8. The $f_2$ values also display a bi-modal distribution around a high and a low value; the center of this lower set of values for $k = 2$ is significantly greater than 0. In the first or '2-core dominant' state, $k = 2$ is the largest core, making up the greatest fraction of particles, surrounded by 1-shell particles, with no 3-core particles present.

In the '3-core dominant' state, the 3-core emerges within the LCC and consists of the greatest fraction of the particles, surrounded by a smaller fraction of 2-shell particles, and an even smaller fraction of 1-shell particles. With increasing $\dot{\gamma}$, the percentage of time the 3-core state is dominant increases, while the frequency of the 2-core dominant state decreases. At large $\dot{\gamma}$, where the stress response has saturated, there is a notable difference in the $k$-core and $k$-shell statistics between CST and DST. In CST for $\phi = 0.54$, the 2-core dominant state is observed even at the highest $\dot{\gamma}$ simulated. On the other hand, for DST at $\phi \geq 0.55$, the 2-core dominant state is no longer observed at the highest shear rate; for any $\phi$ that exhibits DST, we find that there exists a value of $\dot{\gamma}$ above which the 2-core dominant state is not observed. To make this point clearly, in Fig. 13 the temporal distribution of $k$-core particles is displayed for $\phi = 0.54$ and $\phi = 0.55$ at the highest shear rates studied, $\dot{\gamma} = 0.13$ and $\dot{\gamma} = 0.09$, respectively. For $\phi = 0.54$ at $\dot{\gamma} = 0.13$, Fig. 13a shows $k = 2$ to be bimodally distributed. The peak with the highest $f_2$ indicates the presence of the 2-core dominant state, whereas the lower $f_2$ or the 2-shell peak combined with the 3-core distribution represents the 3-core dominant state. In Fig. 13b, for $\phi = 0.55$ at $\dot{\gamma} = 0.09$, the $k = 2$ distribution is no longer bimodal, and the 3-cores are dominant, i.e. have the larger $f_k$. This rather striking change associated with a slight variation of $\phi$ is roughly coincident with the boundary between CST and DST.

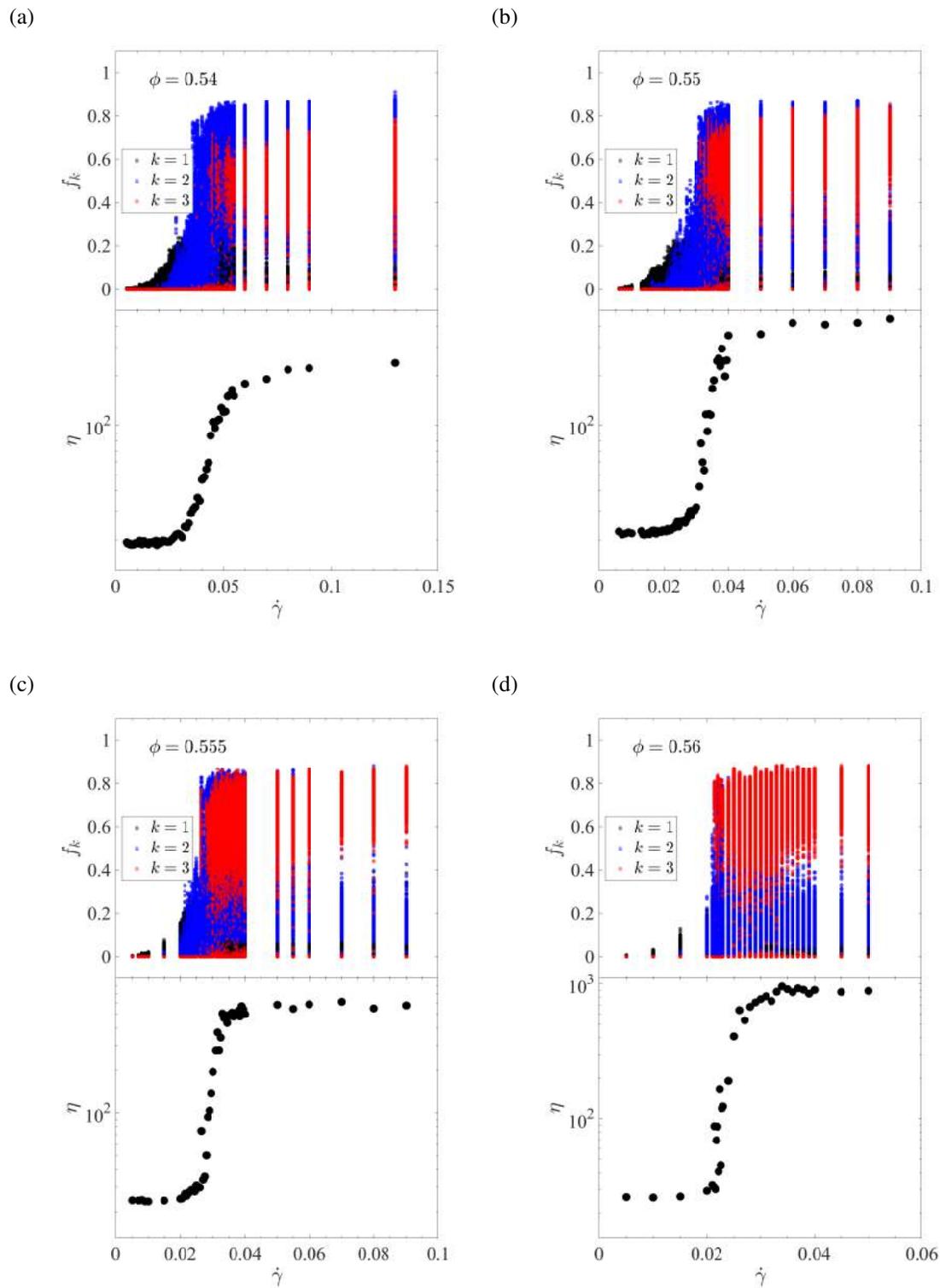

FIG. 12: The fraction of total particles in *k*-cores and *k*-shells of the largest connected component of the frictional contact network as a function of shear rate, $\dot{\gamma}$ plotted along with suspension viscosity for $\phi = 0.54, 0.55, 0.555$ and $0.56$.

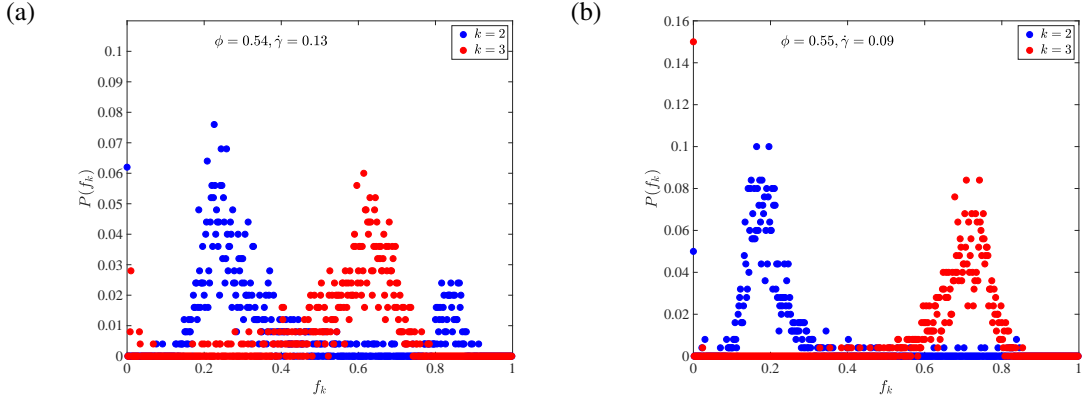

FIG. 13: The temporal distribution of $f_k$ values for $k = 2$ and $k = 3$, (a) for $\phi = 0.54$, $\dot{\gamma} = 0.13$, and (b) $\phi = 0.55$, $\dot{\gamma} = 0.09$. These are the highest shear rate values simulated for each of these volume fractions, as can be seen in Fig. 12.

Recall that Fig. 5 showed that the instantaneous mean degree, $Z$, is a monotonic function of suspension stress at each $\phi$. Earlier, in section III B, we established that the emergence of the GCC is associated with $Z_c \doteq 0.28$. It is natural to extend this line of investigation to the relationship of $Z$ to the emergence and size of the $k$-cores with $k \geq 2$. To this end, we plot the $f_k$ values as a function of $Z$ in Fig. 14. In order to reduce the noise and to differentiate more clearly between the different volume fractions, the $f_k$ values corresponding to the same value of $Z$ were averaged. Similar to the random graph theory, 1-cores emerge continuously and reach their maximum value roughly coincident with the continuous emergence of 2-cores, which rapidly become the maximum core. While $f_1$ decreases, $f_2$ increases until mean degree $Z > 3$.

For $Z \approx 3$, but with some dependence on $\phi$, we observe the emergence of the 3-cores, with this occurring almost discontinuously. We observe a rapid jump from zero to $f_3 \geq 0.3$ for the large majority of samples, with just a few smaller values of $f_3$ seen. This finding is quite similar to the predictions for emergence of the 3-core in a random graph.[44] Note also that at the emergence of the 3-core, $f_2$ drops sharply, as a result of 3-cores emerging from existing 2-cores.

The emergence values of the mean contact number for each $k$ are denoted as $Z_{c,k}$. The $f_1$ curve begins at $Z_{c,1} = 0$, as the first frictional contact established immediately gives birth to a 1-core. The 2-core emergence at $Z_{c,2} \approx 0.28$ can be seen to correspond closely to the maximum of $f_1$, indicating that 2-cores grow by the coalescence of 1-cores. The value of $Z_{c,2}$ is equivalent to the value of $Z_c$ established in section III B, since the birth of the 2-core is found to correspond to the birth of a giant connected component. This indicates that the first 2-cores are large, and in fact most are percolated across the simulated domain, thus connecting one image of the simulation unit cell to the next.

Fig. 14 shows that $f_1$ and $f_2$ collapse on the same curve regardless of the volume fraction for mean degree below the value for the emergence of the 3-core at $Z_{c,3}$. For $Z_{c,3}$, significant differences based on the volume fraction are seen, with $Z_{c,3}$ decreasing as $\phi$ increases. It is observed that with increase of $Z$, driven by the increase of shear rate, 1-cores develop and increase in both number and size; this is akin to nucleation and growth, but since it is with respect to shear rate, not time, this should not be taken as the actual dynamics, as the objects form and disappear in $O(1)$ strain. Once the density of these 1-cores is large enough, they begin to merge or 'coalesce' into 2-cores. By contrast, a 3-core forms suddenly as a giant component from a system-spanning 2-core.

Following this set of observations, we seek to determine whether there is a signature of the emergence of the 3-core in the stress response. We calculate the Z-averaged values of both the fraction of 3-cores ($f_3$) and the shear rate ($\dot{\gamma}$), using the fact that $Z$ is a single-valued monotonic function of $\bar{\sigma}$. The plots of $f_3^Z$ and $\dot{\gamma}^Z$ as functions of $Z$ in Fig. 15 suggest that for each $\phi$, $f_3^Z$ becomes non-zero at a value of $Z$ corresponding to the inflection point of $\dot{\gamma}^Z$, i.e. where $\partial^2 \dot{\gamma}^Z / \partial Z^2 = 0$. The data are rather scattered at $\phi = 0.56$, a condition well into the DST regime, where the suspension is able to exhibit two widely-separated stresses (and hence widely-separated $Z$) at a given shear rate, as shown in Fig. 4.

In the stress response of shear-thickening suspensions, there is a bifurcation[46] that may be interpreted as an apparent critical point that distinguishes between the CST and the DST regimes.[54] This point is found on the flow curve of the lowest $\phi$ for which we observe $\partial \sigma / \partial \dot{\gamma} \to \infty$ (see Fig. 4). The quantity $\partial \sigma / \partial \dot{\gamma} \equiv \chi_\sigma$ may be considered as a stress susceptibility and the fluctuations expected at a divergent susceptibility are addressed in recent work.[54] A striking and important point is that the emergence of the 3-core is coincident with the maximum of this stress susceptibility for each $\phi$ studied. This is illustrated in Fig. 16, where $\chi_\sigma$ and $f_3$ are presented as functions of $\dot{\gamma}$.

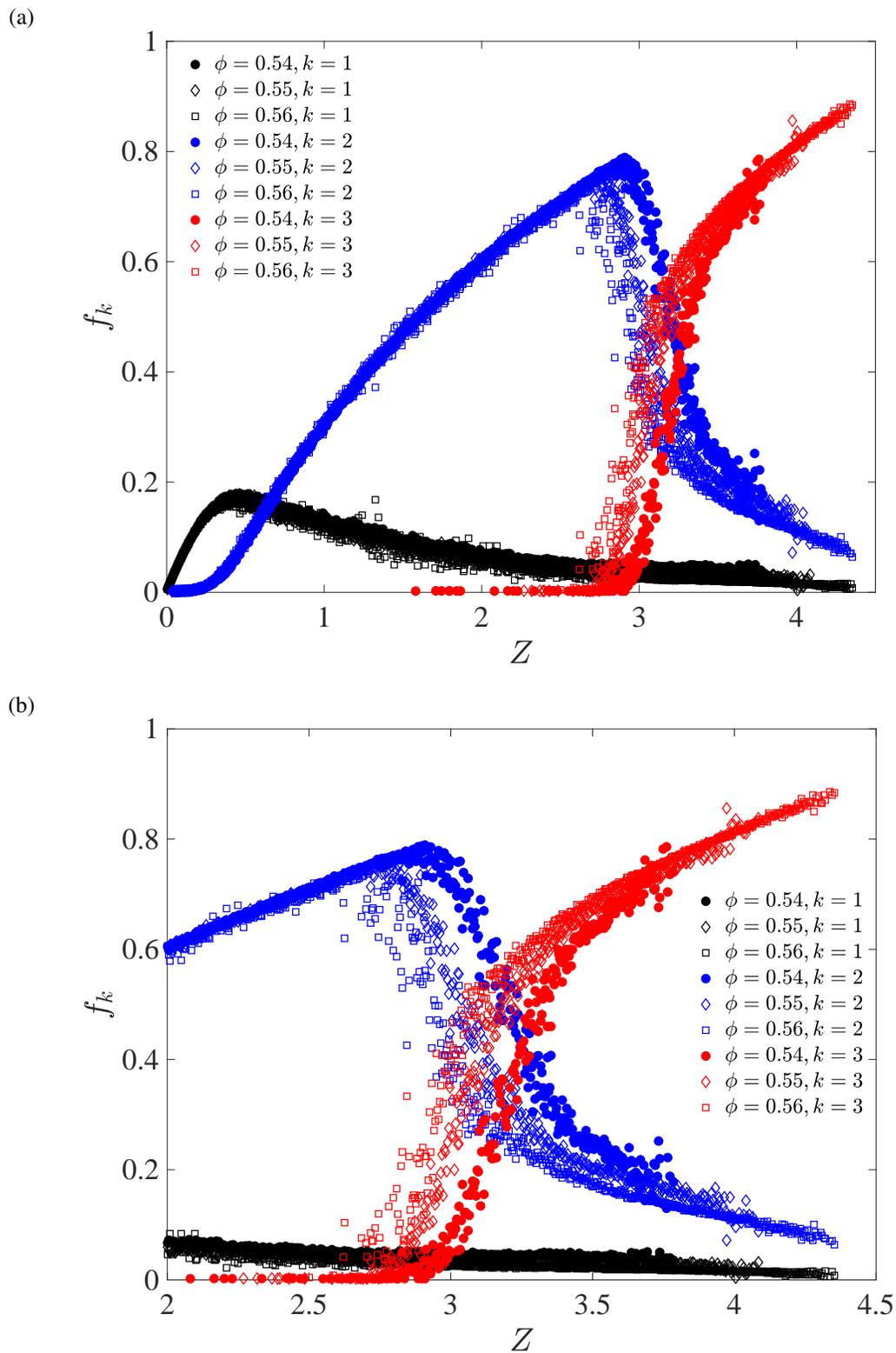

FIG. 14: The $Z$ averaged $k$-shell occupancy, measured as a fraction of total number of particles, $f_k$, plotted as a function of the average frictional network degree, $Z$, for $\phi = 0.54, 0.55$, and $0.56$, across all applied shear rates.

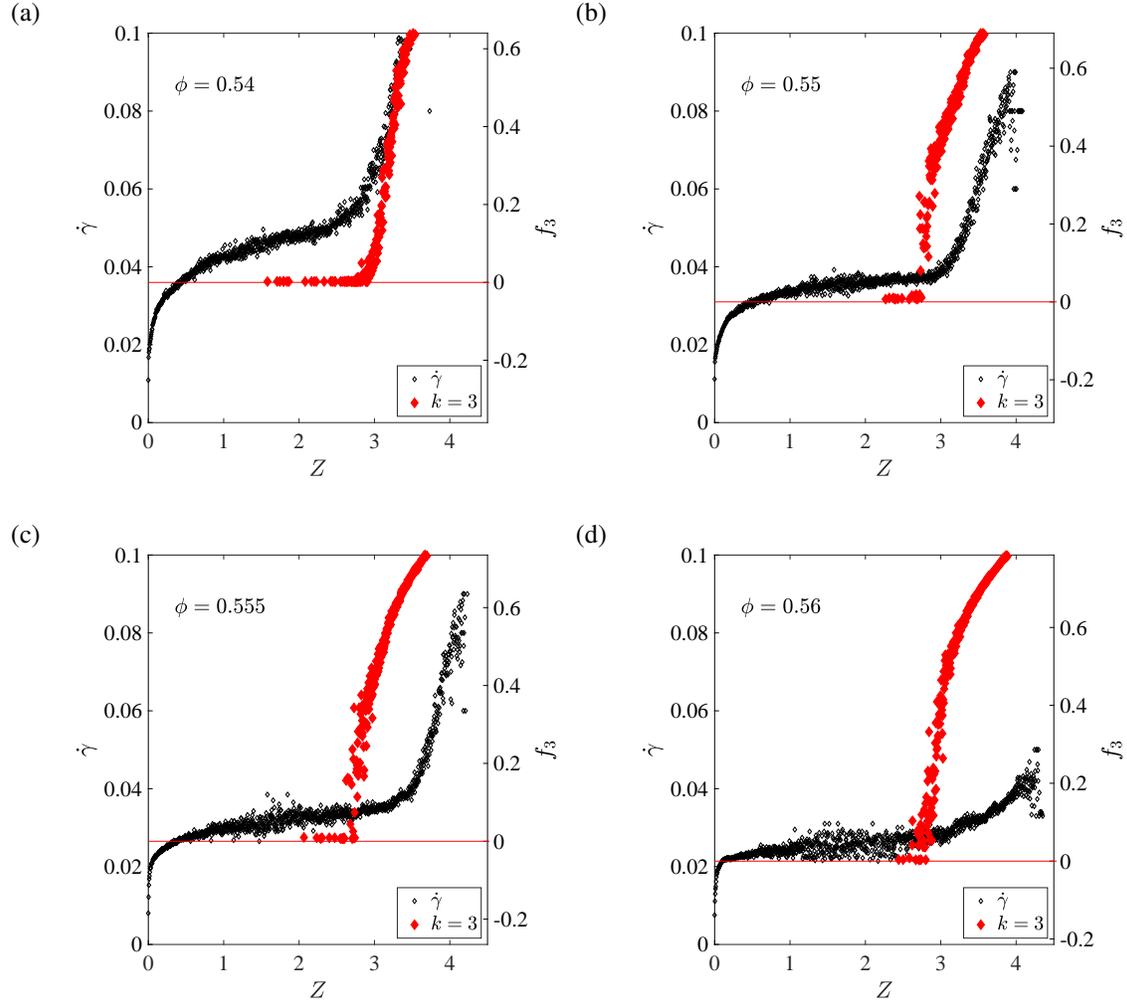

FIG. 15: The $Z$-averaged 3-core occupancy, $f_3$, as a function of mean degree, plotted along with $Z$-averaged shear rate, $\dot{\gamma}$, for $\phi = 0.54, 0.55, 0.555$, and $0.56$. Note that we do not introduce a new symbol, but these are values of $f_3$ and $\dot{\gamma}$ conditioned on the instantaneous mean value of $Z$.

## A. Stress Analysis by $k$-core

Our aim here is to establish a connection between the $k$-cores and the stress in the suspension. The fraction of particles with coreness $k$, $f_k$, is determined at each sampling point. We plot the distribution of all the values of $f_k$ observed, $P(f_k)$, for every $(\phi, \dot{\gamma})$ pair. Next, we investigate the relationship of the coreness of the particle to its stress contribution. For this purpose, we sum the shear stress contribution of the total force moment (hydrodynamic and contact) of all the particles with coreness $k$, and then normalize this quantity by the fraction of the particles with coreness $k$, $f_k = N_k/N$:

$$\bar{\sigma}^k = \frac{1}{f_k} \sum_{i=1}^{N_k} \sigma_i^k. \qquad (2)$$

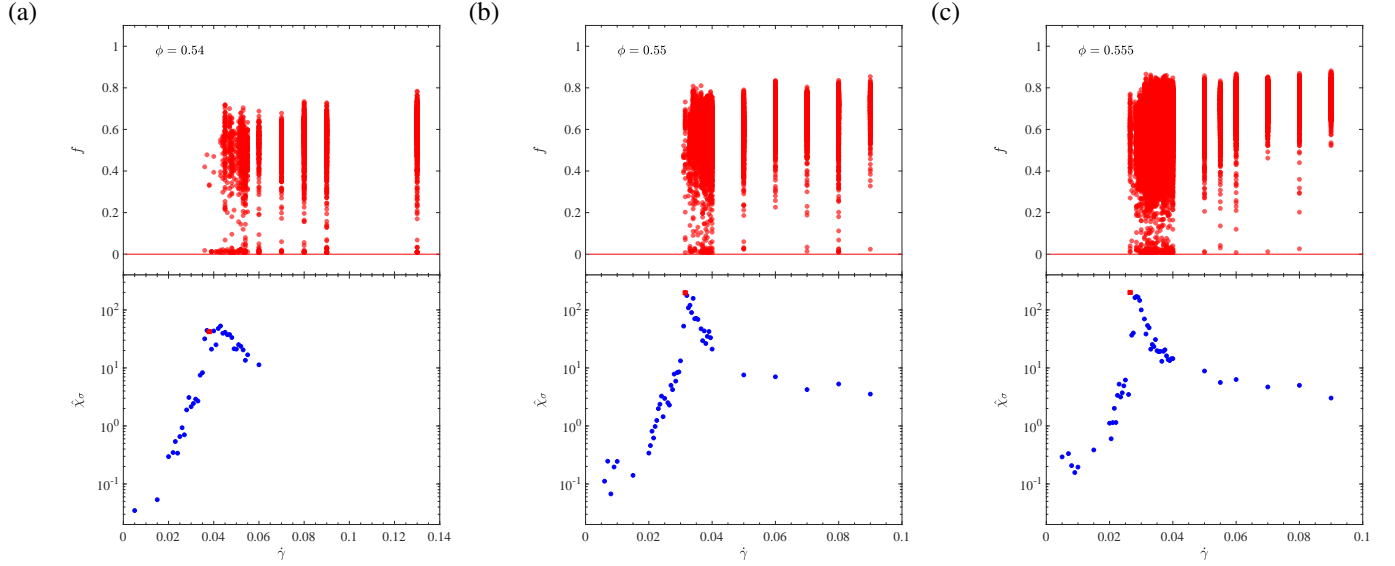

FIG. 16: Fraction of 3-cores (top) and stress susceptibility (bottom) as a function of shear rate, for (a) $\phi = 0.54$, (b) $\phi = 0.55$ and (c) $\phi = 0.555$. The red point in each stress susceptibility plot at bottom corresponds to the first appearance of a 3-core.

This quantity is thus, dimensionally, stress times volume. For each $k$, we normalize this quantity by the total strain units simulated. The distribution of $\bar{\sigma}^k$, $P(\bar{\sigma}^k)$ at each $\dot{\gamma}$ and $\phi$ studied is plotted in Fig. 17. The integral of $P(f_k)$ gives the fraction of time a $k$-core is observed. If a certain core is present throughout the simulation, the integral of $P(f_k)$ is unity.

For all $\phi$, the core-averaged stress distributions as $\dot{\gamma} \to 0$ are similar. Initially particles with purely hydrodynamic interactions ($k = 0$) dominate the stress state, so that the total stress and $k = 0$ contribution are nearly identical. The frictional contacts at this stage are infrequent. When they exist they form small 1-core clusters, and hence the $k = 1$ distribution is noisy, with its integral relatively small compared to the total and the $k = 0$ distributions. As expected, the mean of the $k = 1$ distribution is higher, but within an order of magnitude of the total. As $\dot{\gamma}$ increases, the $k = 1$ distribution becomes more definite as the frictional contacts become more frequent, with a wider

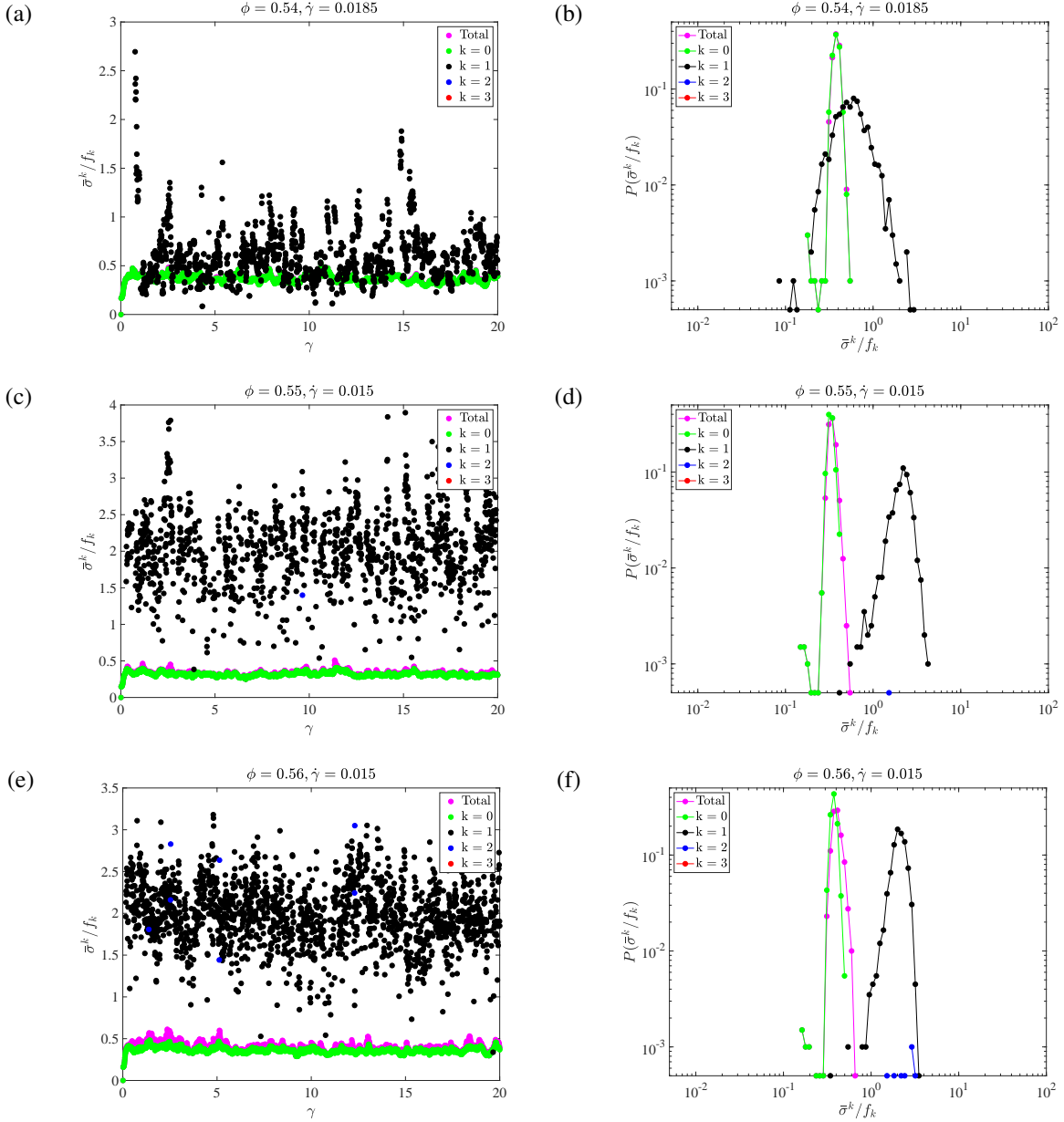

FIG. 17: Core-averaged stress as a function of strain (left) and its distribution (right) at the onset of shear thickening (a, b) $\phi = 0.54$ and $\dot{\gamma} = 0.0185$, (c, d) $\phi = 0.55$ and $\dot{\gamma} = 0.015$, (e, f) $\phi = 0.56$ and $\dot{\gamma} = 0.015$. Note that no 3-cores (red points) are present, and only very few 2-cores (blue points).

distribution compared to both the hydrodynamic and the total distributions as seen in Fig. 17.

The $k = 2$ distribution is the next to appear. It is initially noisy, and yields a relatively small integral. A more definite distribution and a larger integral is found as the shear rate increases and the 2-cores become more frequent. Unlike $k = 1$, the evolution and the value of the mean stress for $k=2$ is different for CST ($\phi = 0.54$) and DST ($\phi \geq 0.55$) conditions. Specifically, under CST conditions at $\phi = 0.54$, the mean value of $\sigma$ for $k = 2$ is similar to or slightly less than the mean value of the $k = 1$ distribution, while for $\phi = 0.55$ and $\phi = 0.56$ (at onset and well into DST, respectively), the $k = 2$ mean stress level is significantly larger than that for $k < 2$.

At shear rates just below that yielding the maximum stress susceptibility, i.e. just below the appearance of the 3-core, hydrodynamic ($k = 0$), $k = 1$ and $k = 2$ distributions are all well-defined and integrate to 1, since they are present at all times during the simulation; see Fig. 18. For $\phi = 0.54$, the $k = 1$ and $k = 2$ distributions almost overlap, and even more remarkably, the distributions for all $k$ and hence the box-averaged stress distribution have the same form: this is asymmetric with a positive skew, where the median is at the lower $\sigma$ half of the distribution, and with a decaying tail for large $\sigma$. In CST, the core-averaged stresses for 1- and 2-core particles are similar.

For $\phi \geq 0.55$, the situation is significantly different than just described for $\phi = 0.54$. At shear rates just below the appearance of the 3-core, the core-averaged distributions are different in form, and their mean values are distinctly different, increasing from $k = 0$ to $k = 1$ to $k = 2$. The $k = 0$ and $k = 1$ distributions both have negative skew, whereas $k = 2$ has a positive skew. The system-averaged stress, with a positive skew, has the widest distribution, covering the range of all the core-averaged stress distributions. This demonstrates that for $\phi$ yielding DST, the stress state of a particle and the core to which it belongs are meaningfully correlated and differ with $k$; particles with coreness $k = 2$ are more likely to be found at a higher stress than particles with $k = 1$. As a result, the suspension stress has a wider range of values; at a given instant, the stress assumes a value closest to the core-averaged stress of the dominant core.

The emergence of the 3-core is found to occur at shear rate corresponding to the maximum stress susceptibility for all $\phi$. Recall that, unlike 1- and 2-cores, the 3-core transition is discontinuous, and the 3-core emerges as a single system-spanning cluster of a finite fraction of the system size. Therefore, large temporal fluctuations in the formation of the 3-core cluster, such that it forms and collapses rapidly, translate to stress fluctuations $\sim f_3 \bar{\sigma}^3$. With this in mind, we consider $\bar{\sigma}^3$ in Fig. 19. For CST, at $\phi = 0.54$ and $\dot{\gamma} = 0.044$, the $\bar{\sigma}^3$ distribution overlaps with other cores and non-frictional particles. Thus, in CST, on average a particle with $k = 3$ contributes to the stress at a level similar to that of a $k = 2$ or a $k = 1$ particle. As a result, even though we have large fluctuations in the frictional network structure as particles form and break out of a 3-core, this does not lead to large stress fluctuations. The system averaged stress distribution, $P(\bar{\sigma})$, despite developing a wide distribution during the shear-thickening transition, does not develop a two-peak

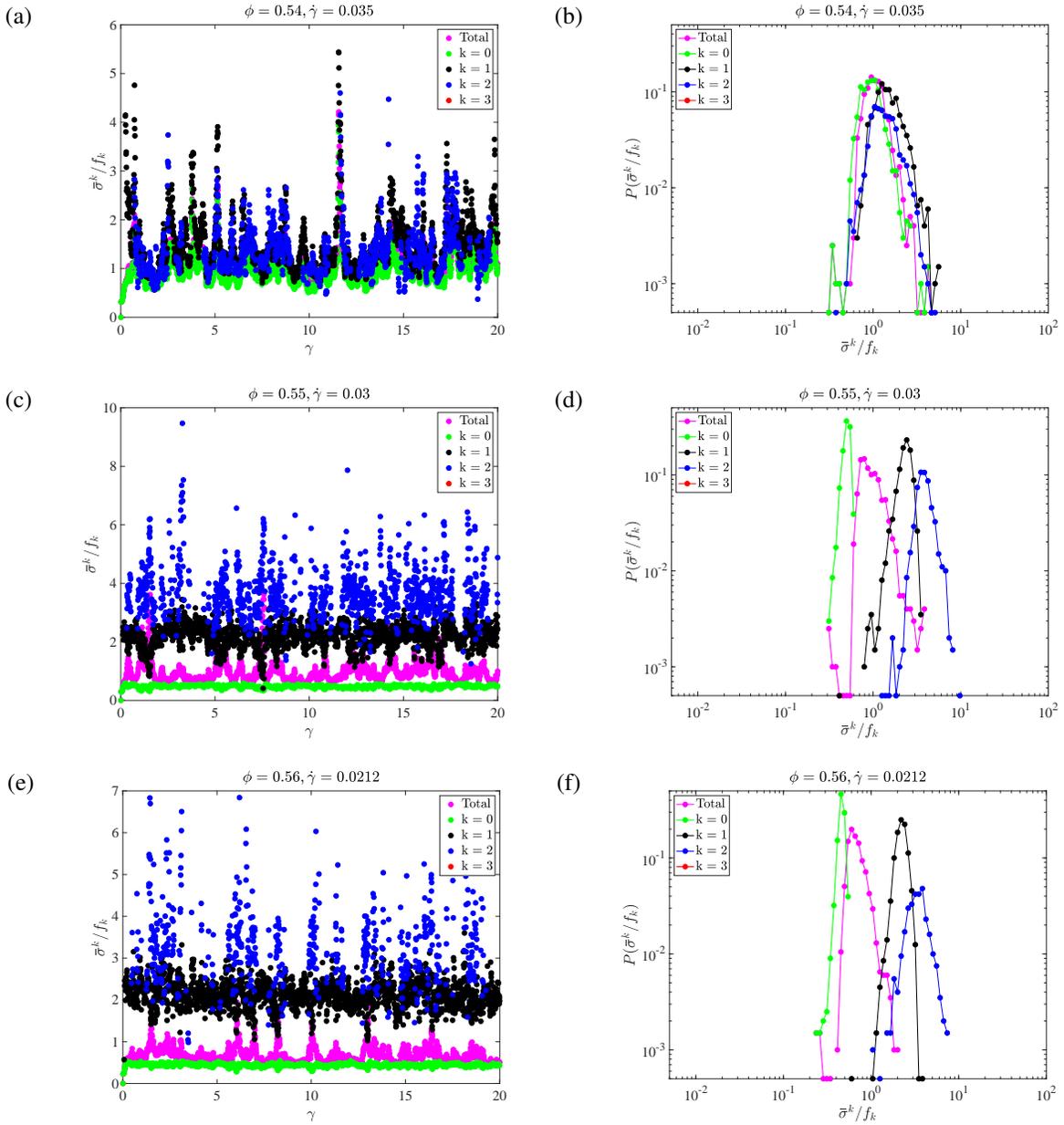

FIG. 18: Core-averaged stress as a function of strain (left) and its distribution (right) at the beginning of strong upturn in viscosity (a, b) $\phi = 0.54$ and $\dot{\gamma} = 0.035$, (c, d) $\phi = 0.55$ and $\dot{\gamma} = 0.03$, (e, f) $\phi = 0.56$ and $\dot{\gamma} = 0.0212$. Note that no 3-cores (red points) are present.

distribution indicative of exchanging stress states.

For $\phi = 0.55$, the emergence of the 3-core at $\dot{\gamma} = 0.0315$ represents a large fraction of particles ($f_3 = 0.3$ to $0.7$) when it is present but this is only a small fraction of the time initially, and remarkably contributes a core-averaged stress an order of magnitude larger than the $k = 2$ particles. The $\bar{\sigma}^3$ distribution, unlike in CST conditions, is distinct, with relatively small range and a mean stress larger by roughly one and two orders of magnitude than the $k = 2$ and $k = 0$ mean stress levels, respectively. The implication is that the effect of the 3-core on the total stress state of the suspension is large for $\phi = 0.55$. The system-averaged stress distribution, $P(\bar{\sigma})$, assumes a much wider range, as the 3-core becomes more persistent and is present for a larger fraction of the time. As shear rate increases to $\dot{\gamma} = 0.033$, we observe a second peak forming in the total stress distribution, one peak between $k = 0$ and $k = 1$, representing the low stress state, and the other between $k = 2$ and $k = 3$, representing the high stress state.

For $\phi = 0.56$, at $\dot{\gamma} = 0.0218$ we have the $k = 3$ distribution entering. This is the lowest value of the shear rate where we observe the system-average stress splitting into two sub-distributions, representing the temporal exchange between the low- and high-stress states. The high-stress part of the distribution is nearly overlapping with the 3-core stress distribution.

## V. CONCLUDING REMARKS

We have investigated how the force network in shear-thickening suspensions develops with increasing shear rate in the volume fraction region spanning the transition from continuous to discontinuous shear thickening. This has been done starting with basic measures and progressing to a specific type of sub-graphs, known as $k$-cores. The known correlation of the increase of bulk stress with the development of the contact force network between frictional particles has been elucidated, with $k$-core analysis providing both a clear correlation of these network structures with the increase of the apparent viscosity of the suspension, and also showing a sharp distinction between CST and DST.

The development of network structure is characterized at its most basic level by the mean frictional contact number, $Z$, which is found to be a strictly increasing function of the stress at a given $\phi$. This one-to-one relationship between contact number and stress for a given suspension provides a basis for describing the behavior through the network features; further, it indicates that discontinuities in $\sigma$ in DST or $\partial \sigma / \partial \dot{\gamma} \to \infty$ at the lowest $\phi$ for DST (the critical volume fraction of $\phi_C \approx 0.55$) are associated with discontinuities in the mean contact number. This is clearly related to the 'fraction of frictional contacts' $f(\sigma)$ (a local notation, not to be confused with other meanings of $f$ in this paper) from the work of Wyart & Cates,[46] which also undergoes a discontinuous change in DST when considered as a function of shear rate rather than stress.

A more detailed consideration of the contacts through the degree distribution, i.e. the probabil-ity distribution of the number of contacts a particle is expected to have for varying $\dot{\gamma}$ and $\phi$, is not greatly informative in relation to the change from CST to DST, at least to the level considered here. This shows the development of progressively more particles at nonzero contact numbers, with the distribution peaked at $Z_k = 3$-$4$ contacts for large shear rate, with this peak at larger $Z_k$ for larger $\phi$. Perhaps more important is that the maximum encountered (only occasionally) is eight contacts per particle; thus all $Z_k$ values are significantly below the maximum of 12 for dodecahedral arrangement in a dense packing.

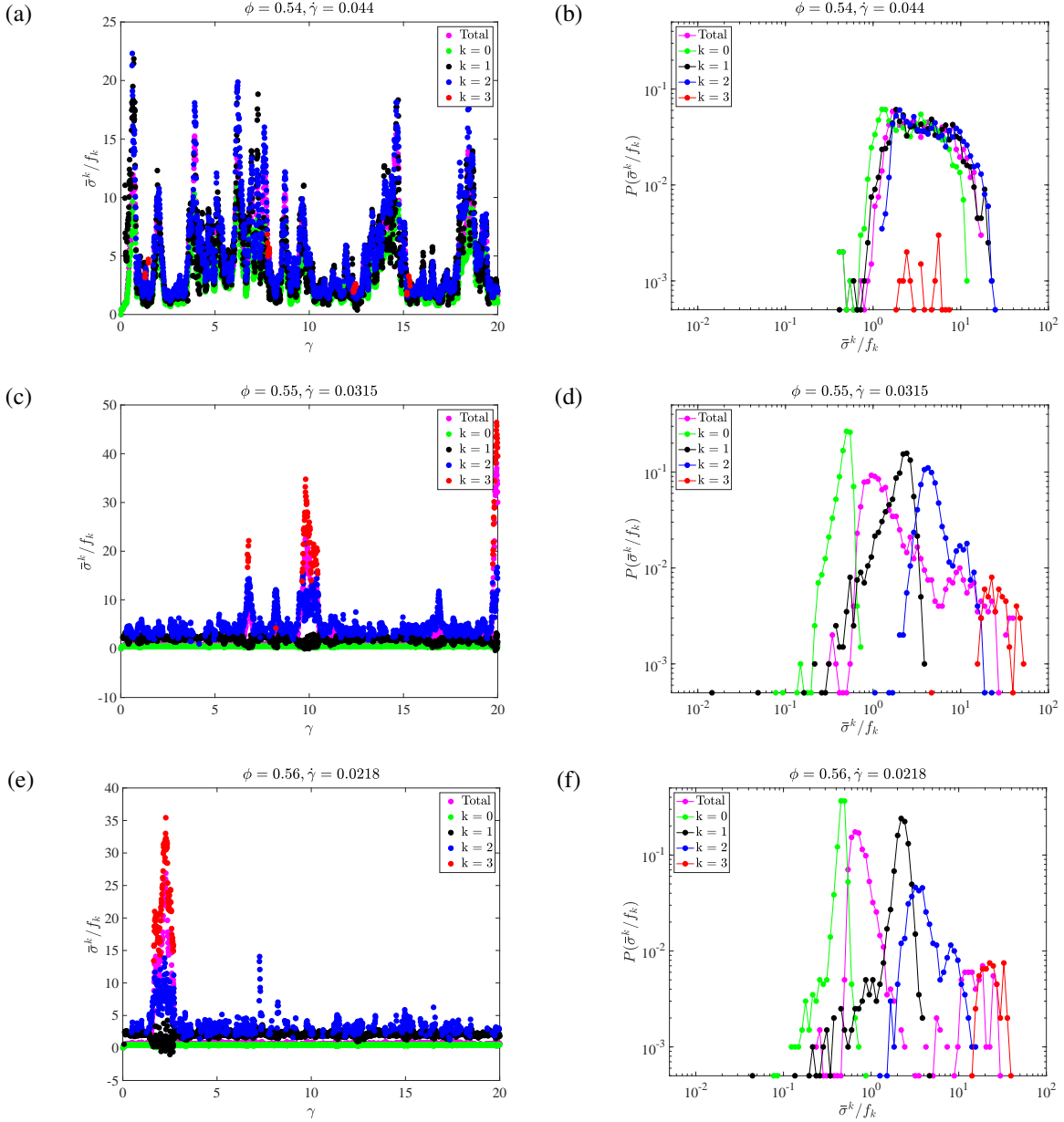

FIG. 19: Core-averaged stress as a function of strain (left) and its distribution (right) at peak in stress susceptibility (a, b) $\phi = 0.54$ and $\dot{\gamma} = 0.044$, (c, d) $\phi = 0.55$ and (e, f) $\dot{\gamma} = 0.0315$, $\phi = 0.56$ and $\dot{\gamma} = 0.0218$. Note that 3-cores are present and particles in them have significantly larger stress than in other $k$-cores.

The need to consider longer range connectivity to explain the rheological transitions in shear thickening brings percolation as a candidate. In fact, percolation has been suggested to be the basis for onset of DST.[53] However, we have applied the Molloy-Reed criterion[65] to show that percolation occurs in both CST and DST and thus is not a discriminating measure. An interesting result that arises is that the mean contact number at percolation for $\phi = 0.54$ (CST) and 0.56 (DST) is both the same within our sampling uncertainty, $Z_c \approx 0.28$, and strikingly below the value of $Z_{c,random} = 1$ for random graph percolation.

Measures of connectivity between the particles based on $k$-core analysis show quite clear relationships with three aspects of the suspension rheology. The $k$-cores are sub-graphs of the complete frictionally contacting network, all nodes (particles) having $k$ or more contacts with other nodes in the sub-graph. The first relation with the rheological response we note is that the appearance of the 3-core for a given $\phi$ as $\dot\gamma$ increases coincides (as shown by Fig. 15 and Fig. 16) with the peak in the stress susceptibility defined[54] as $\chi_\sigma = \partial\sigma/\partial\dot\gamma$; this is true under conditions of both CST and DST in the range of $\phi$ studied here. Although the suspension connectivity is certainly not random, the 3-core emerges as a giant connected component of a significant fraction ($\approx 0.3$) of the total number of particles in the network, similar to expectations based on random graph theory.[44] A second feature that indicates the $k$-core's relation to the mechanical response is the difference in the stress contribution by particles in 3-cores at the onset of DST relative to CST conditions, as detailed in section IV A. In CST, a particle in the 3-core makes a stress contribution larger than but still comparable to that from a particle in the 2-core (with statistical distributions of 2- and 3-cores overlapping), while for the DST condition, the 3-core particles at emergence of this structure have much larger stress than smaller-$k$ particles. A third relationship is that we find under conditions of CST (here at $\phi = 0.54$), considering the large-stress limit where the network structure and rheological properties have saturated, the 2-core and 3-cores exchange with time (or strain) as the dominant structures in the sense of containing the larger fraction of the total particles. By contrast, even at the lowest $\phi$ displaying DST, $\phi = 0.55$, the 3-core is always dominant in the large-stress limit.

These investigations have shown that $k$-core analysis provides insight to features of the network formed by frictional contacts that define the stress response. The natural next stage of study is to explore the mechanical basis for this influence, e.g. its potential relationship to slowly relaxing regions or transient locally rigidified regions within the flowing suspension.

ACKNOWLEDGEMENTS

This work was supported by a collaborative grant, NSF CBET-1916877 (BC) and NSF CBET-1916879 (JFM), and by NSF DMR-1945909 (HAM). Computations in this work were supported, in part, under the NSF grants CNS-0958379, CNS-0855217, and ACI-1126113 to the City University of New York High Performance Computing Center at the College of Staten Island.